\documentclass{aa}

\usepackage{graphicx}
\usepackage{amsmath}   
\usepackage{txfonts}
\usepackage{booktabs}
\usepackage{xcolor}
\usepackage{xspace}
\usepackage{tikz}
\usepackage[version=4]{mhchem}
\usetikzlibrary{shapes.geometric,arrows.meta,positioning,fit,backgrounds}
\usepackage[colorlinks=true,linkcolor=blue,citecolor=blue,urlcolor=blue]{hyperref}

\newcommand{\bear}{BeAR\xspace}
\newcommand{\heliosr}{\texttt{Helios-r2}\xspace}
\newcommand{\poseidon}{\texttt{POSEIDON}\xspace}
\newcommand{\chimera}{\texttt{CHIMERA}\xspace}
\newcommand{\multinest}{\texttt{MultiNest}\xspace}
\newcommand{\nautilus}{\texttt{Nautilus}\xspace}
\newcommand{\fastchem}{\texttt{FastChem}\xspace}
\newcommand{\heliosk}{\texttt{Helios-k}\xspace}
\newcommand{\sysrem}{\textsc{SysRem}\xspace}
\newcommand{\Kp}{K_\mathrm{p}}
\newcommand{\Vsys}{V_\mathrm{sys}}
\newcommand{\IP}{(\mathbf{I}-\mathbf{P})}

\begin{document}

\title{BeAR: The Bern Atmospheric Retrieval code}
\subtitle{A GPU-accelerated framework for low-resolution, high-resolution, and joint retrievals of exoplanet atmospheres}

\titlerunning{BeAR: The Bern Atmospheric Retrieval code}
\authorrunning{E. Wang et al.}

\author{E. Wang\inst{1}
  \and D. Kitzmann\inst{1,2}
  \and H. J. Hoeijmakers\inst{3}
  \and J. P. Wardenier\inst{1}
  \and H. Lecoq-Molinos\inst{1}
  }

\institute{
  Space Research and Planetary Sciences, Physics Institute, University of Bern,
  Gesellschaftsstrasse 6, 3012 Bern, Switzerland\\
  \email{esther.wang@unibe.ch}
  \and
  Center for Space and Habitability, University of Bern,
  Gesellschaftsstrasse 6, 3012 Bern, Switzerland
  \and
  Lund Observatory, Division of Astrophysics, Department of Physics,
  Lund University, Box 43, 221 00 Lund, Sweden
  }

  \date{}

\abstract
{Joint retrievals of ground-based high-resolution and space-based low-resolution exoplanet spectra are becoming more common. High-resolution results depend on analysis choices without a low-resolution counterpart, such as the removal of telluric and stellar lines and the mapping from cross-correlation to likelihood, and therefore need to be tested with independent codes. Yet few open-source high-resolution retrieval codes exist, and runtimes of hours to days make such tests expensive.}
{We present \bear, the Bern Atmospheric Retrieval code and successor of \heliosr, and its new high-resolution cross-correlation framework for emission and transmission spectra. \bear performs joint retrievals by adding the log-likelihoods of high- and low-resolution data computed from the same model atmosphere.}
{The framework implements established cross-correlation log-likelihoods, with the line-depth scaling either fitted analytically or sampled as a free parameter, their per-pixel generalization, fast model filtering, and model re-injection adapted to this filtering. The forward model and likelihood run entirely on a graphics processing unit (GPU). One forward model takes about 1 ms on an NVIDIA GH200, allowing millions of models per minute on a GPU cluster.}
{In separate retrievals of Gemini South/IGRINS and JWST/NIRSpec dayside spectra of the hot Jupiter WASP-77A\,b, \bear{} reproduces published abundances, C/O ratios, and metallicities to within $1.5\sigma$, the preference for $^{13}$CO, and the thermal structure and line contrast to within $1\sigma$. The joint retrieval confirms a sub-solar metallicity and a near-solar C/O ratio. A full IGRINS retrieval with 17 free parameters takes 27 to 42 minutes on a single NVIDIA GeForce RTX 5080 or about 14 minutes on eight RTX 4090 GPUs. Injection--recovery tests show that the posteriors underestimate the uncertainties of abundance ratios by about a factor of two.}
{\bear is an open-source, GPU-accelerated code for low-resolution, high-resolution, and joint retrievals that makes tests requiring many retrievals practical. It is available at \url{https://github.com/NewStrangeWorlds/BeAR}.}

\keywords{
  techniques: spectroscopic --
  planets and satellites: atmospheres -- 
  methods: numerical}

\maketitle
\nolinenumbers

\section{Introduction}
\label{sec:intro}

The atmosphere of an exoplanet can be observed at widely different spectral resolving powers, and each regime retains different information. At low resolution, with $R=\lambda/\Delta\lambda$ up to a few thousand, molecular bands are sampled as a whole together with the continuum between them, which constrains the absolute abundances, the cloud opacity, and, in emission, the temperature gradient \citep{Benneke2012,Line2013}. JWST now covers the bands of \ce{H2O}, CO, and \ce{CO2} within a single observation \citep[e.g.,][]{August2023}, so the metallicity and the C/O ratio, which trace planet formation \citep{Oberg2011}, can in principle be constrained from one spectrum, although different instruments have yielded different metallicities for the same planet \citep[e.g.,][]{Sing2024,Welbanks2024}. Within each band, however, the lines remain blended, so species with overlapping bands and the temperature structure can partly compensate for one another \citep{Brogi2017}.

High-resolution spectroscopy ($R\gtrsim25\,000$) resolves these bands into individual lines \citep[for reviews, see][]{Birkby2018,Snellen2025}. The changing Doppler shift of the planet's lines during the observation separates them from the nearly stationary telluric and stellar lines, and cross-correlation with a model template combines the many lines that individually lie below the noise \citep{Snellen2010,Brogi2012,Birkby2013}. The distinct line patterns separate species with overlapping bands \citep{Brogi2017,Line2021}, and lines of different strength probe different pressures \citep{Brogi2019}. However, the normalization and detrending steps used to remove the telluric and stellar lines also remove the broadband shape of the planetary spectrum, so high-resolution data alone constrain abundance ratios more tightly than absolute abundances \citep{Brogi2019,Gibson2022}.

Measuring abundances and temperature structures from such data can be done with a Bayesian atmospheric retrieval using a parametric forward model \citep{Madhusudhan2009,Line2013,MacDonald2023}, and at high resolution such retrievals need two additional ingredients. The first is a mapping from cross-correlation to likelihood \citep{Zucker2003,Brogi2019}, which \citet{Gibson2020} generalized to per-pixel uncertainties. The second is a consistent treatment of the detrending. \sysrem{} \citep{Tamuz2005,Birkby2013} and principal component analysis \citep[PCA;][]{deKok2013} remove part of the planetary signal together with the telluric and stellar lines, so every trial model must be distorted in the same way. Repeating the detrending for each model \citep{Brogi2019,Line2021,Pelletier2021} is slow and can dominate the cost of a retrieval. For example, a retrieval of the IGRINS data of WASP-77A\,b took about three days on 24 central processing unit (CPU) cores and one graphics processing unit (GPU) \citep{Line2021}. \citet{Gibson2022} replaced this step by filtering each model with a fixed linear projection derived once from the data, which they found to be orders of magnitude faster than repeating \sysrem{} and substantially faster than repeating PCA for every model. Such retrievals have measured the C/O ratios and metallicities of hot Jupiters \citep{Line2021,Brogi2023,Smith2024b}, refractory-to-volatile ratios \citep{Pelletier2025}, and relative metal abundances from transmission spectra \citep{Gibson2022,Maguire2023,Gandhi2023}.
 
Fitting high- and low-resolution data with one atmosphere can break degeneracies that neither resolves alone. The low-resolution spectrum retains the absolute flux and, in emission, the flux difference between strongly absorbing bands and the windows between them, which probe different pressures. It therefore constrains the absolute abundance scale and the temperature gradient, whereas resolved lines separate species whose bands overlap and also give access to Doppler shifts caused by the planet's orbital motion, rotation, and winds \citep{Brogi2017,Brogi2019}. \citet{Smith2024} performed the first such joint retrieval with JWST data, combining IGRINS and NIRSpec/G395H dayside spectra of WASP-77A\,b, and obtained tighter constraints than from either instrument alone; earlier joint fits used \textit{Hubble} and \textit{Spitzer} data \citep{Brogi2017,Gandhi2019,Kasper2023,Boucher2023}. As JWST and ground-based spectra accumulate for the same planets, such analyses are likely to become more numerous.
 
The results of high-resolution and joint retrievals, however, depend on analysis choices such as the detrending method, the number of removed principal components, the model filtering, and the form of the likelihood \citep[e.g.,][]{Cabot2019}. \citet{Smith2024} find that their post-eclipse IGRINS constraints depend strongly on the number of removed components, and \citet{Wang2025} find a spurious \ce{NH3} detection that appears with \sysrem{} but not with PCA. At low resolution, such choices are often tested by analyzing the same spectra with several independent codes \citep{Barstow2020,Fisher2024,Schmidt2025}, which at high resolution requires open-source implementations. These have become available only recently, as extensions of \texttt{petitRADTRANS} \citep{Molliere2019,Nasedkin2024,Blain2024joss,Blain2024} and \poseidon{} \citep{MacDonald2017,MacDonald2023joss,Wang2025}, as the stand-alone code \texttt{ATMOSPHERIX} \citep{Klein2024,Debras2024}, and as the \chimera{}-based example code released with \citet{Line2021}. To our knowledge, none of them has yet reproduced a published joint analysis, and runtimes of hours to days per retrieval \citep{Line2021,Wang2025} make tests that require many retrievals, such as injection--recovery tests, expensive.
 
In this paper we present \bear, the Bern Atmospheric Retrieval code, and its extension to high-resolution cross-correlation retrievals. \bear is the open-source successor of \heliosr{} \citep{Kitzmann2020} and has been applied to brown dwarfs \citep{Lueber2022} and to \textit{Hubble} and JWST transmission spectra \citep{Lueber2024,Fisher2024,Lueber2025,Schmidt2025}. The new module implements the likelihoods of \citet{Brogi2019} and \citet{Gibson2020, Gibson2022}, the fast filtering of \citet{Gibson2022}, and a version of the model re-injection of \citet{Line2021} adapted to this filtering, as in \poseidon \citep{Wang2025}. Each observation enters as a separate term of one log-likelihood computed from a common atmospheric model, so a joint fit requires only listing the observations. Because everything except the atmospheric structure and chemistry runs on the GPU, a high-resolution retrieval of the IGRINS data of WASP-77A\,b takes less than an hour on a single consumer GPU and about 14 minutes on eight GPUs.
 
\section{The \bear framework}
\label{sec:bearold}

\bear is an open-source Bayesian retrieval code for the atmospheres of exoplanets and brown dwarfs. It was first developed under the name \heliosr \citep{Kitzmann2020} as part of the \textsc{Helios} family of codes, which also includes the atmospheric model \textsc{Helios} \citep{Malik2017} and the opacity calculator \heliosk{} \citep{GrimmHeng2015,Grimm2021}. It was later renamed \bear to avoid confusion within the growing family of \textsc{Helios} codes. \heliosr was itself a complete rewrite of \textsc{Helios-r} \citep{Lavie2017}, an earlier, Python-based retrieval code that was never made public. \bear currently performs retrievals of emission, transmission, and secondary-eclipse spectra. It accepts photometry and low- to medium-resolution spectra, given either at individual wavelengths or binned into wavelength bands, and, with this work, also high-resolution data. The code is released under the GNU General Public License (GPL-3.0) on GitHub together with its documentation\footnote{\url{https://github.com/NewStrangeWorlds/BeAR}}.
 
The core of \bear is written in C++ and NVIDIA's CUDA language. All computationally heavy steps run on the GPU, namely opacity interpolation, optical depths, radiative transfer, convolution with the instrument profile, binning to the observed wavelengths, and likelihood evaluation. The atmospheric structure, including the equilibrium chemistry with \fastchem{} \citep{Stock2018,Stock2022,Kitzmann2024,Kitzmann2026}, is computed on the CPU (Sect.~\ref{sec:cuda}). The complete code can also run on a pure CPU setup at the cost of a slowdown by a factor of 100 or more.  
 
Bayesian inference uses nested sampling \citep{Skilling2006}, which provides posterior distributions together with the Bayesian evidence for model comparison. Retrievals run from a Python interface with, for example, \multinest \citep{Feroz2008, Feroz2009} via \texttt{PyMultiNest} \citep{Buchner2014}. \bear can also be run with other samplers with a Python interface, such as \texttt{dynesty} \citep{Speagle2020} or \nautilus \citep{Lange2023}. The same Python interface gives direct access to the forward models, so that spectra can be computed for user-defined atmospheric structures without a retrieval. The forward models can also be used with other parameter-space explorers or to compute training sets for machine-learning retrievals, such as the random forest that \citet{Fisher2020} applied to high-resolution cross-correlation data.

\subsection{Forward models}
\label{sec:bear_forward}

The following subsections summarize the components of \bear, each of which the online documentation describes in detail. \bear contains separate forward models for the main observing geometries, all of which describe a one-dimensional atmosphere divided into layers that are equidistant in log-pressure. The transmission model computes the wavelength-dependent transit depth, with the surface gravity, the planet radius at the bottom of the modeled atmosphere, and the stellar radius as general parameters. Optionally, the mean molecular weight or the atmospheric scale height can replace the composition-derived values as free parameters when the background gas is unknown. The secondary-eclipse model computes the occultation depth, with the stellar spectrum supplied as a blackbody, a tabulated spectrum, or a spectrum interpolated from a grid of stellar atmosphere models. The emission model computes the flux $F(\lambda)$ of a self-luminous object scaled by the radius--distance relation of \citet{Kitzmann2020}. Two simpler models, a secondary-eclipse model with a planetary blackbody and a flat line, serve as references for Bayesian model comparison, for example, to test whether the data warrant an atmosphere model at all. The phase-curve model used for high-resolution data is described in Sect.~\ref{sec:highres}.

The original \heliosr code solved the radiative transfer equation with the method of short characteristics \citep{OlsonKunasz1987}, neglecting scattering. This solver remains available in \bear, together with two more advanced ones. A new C++ implementation of the discrete-ordinate solver DISORT \citep{Stamnes1988}, \textsc{DisORT++}\footnote{\url{https://github.com/NewStrangeWorlds/DisORT}}, provides a multi-stream reference solution on the CPU. A C++/CUDA implementation\footnote{\url{https://github.com/NewStrangeWorlds/AddingDoublingRT}} of the adding-doubling method is now also available on both CPU and GPU, providing a general multi-stream radiative transfer solver including multiple scattering. The transmission model integrates the slant optical depths along the limb geometry, with the option of a gravity that varies with altitude as described in \citet{Bower2019}.

\subsection{Temperature profiles}
\label{sec:bear_temperature}

The vertical temperature structure can be described by several parameterizations of increasing flexibility, the simplest of which is an isothermal profile. Two analytic profiles follow from radiative equilibrium, namely Milne's solution for a gray, self-luminous atmosphere with the exact Hopf function \citep{Mihalas1978,Kitzmann2020} and the profile of \citet{Guillot2010} for atmospheres heated by either a collimated stellar beam or isotropic irradiation. The three-layer parametric profile of \citet{Madhusudhan2009}, which describes both monotonic profiles and thermal inversions with six parameters, is also available and is used for the emission retrievals in this work (Sect.~\ref{sec:validation}). 
 
Free profiles can be described by piecewise polynomials on a finite-element discretization in log-pressure \citep{Kitzmann2020}, by cubic B-splines, or by a monotone piecewise-cubic Hermite interpolant \citep[PCHIP;][]{FritschCarlson1980}, which is shape preserving and avoids the oscillations that global splines can introduce between regularly spaced nodes. For these free profiles, the temperatures at the nodes can be sampled either directly or through the bottom temperature and successive ratio factors, which allows monotonic profiles to be enforced through the priors. Finally, a radiative-convective profile joins a convective adiabat in the deep atmosphere to a PCHIP spline above the radiative-convective boundary, with the adiabatic index and the location of the boundary as free parameters.

\subsection{Chemistry}
\label{sec:bear_chemistry}

Chemical abundances can be retrieved freely or computed self-consistently under chemical equilibrium. In the free approach, each species is assigned a constant-with-altitude volume mixing ratio, sampled either with standard log-uniform priors or with centered-log-ratio priors, which treat all species symmetrically when no dominant background gas can be assumed \citep{Benneke2012}. Vertically varying free profiles are available through piecewise polynomials, cubic B-splines, and PCHIP interpolation, analogous to the temperature treatment. A two-level step function based on \citet{Rowland2023} describes a species that condenses within the atmosphere.  
 
For retrievals with equilibrium chemistry, \bear couples directly to the equilibrium chemistry code \fastchem \citep{Stock2018,Stock2022,Kitzmann2024,Kitzmann2026}, with the metallicity factor and, optionally, individual elemental ratios such as C/O or Ti/H as free parameters. Equilibrium condensation and rainout are available through \fastchem. Several chemistry models can be combined in one retrieval, for example, an equilibrium background with a free profile for a species that is expected to be out of equilibrium. 

A background chemistry model can fill the remainder of each layer, either with H$_2$ and He, with a single species such as N$_2$, or with a temperature-dependent H/H$_2$/He equilibrium mixture for atmospheres in which hydrogen is partly dissociated.

\section{High-resolution retrieval in \bear{}}
\label{sec:highres}
 
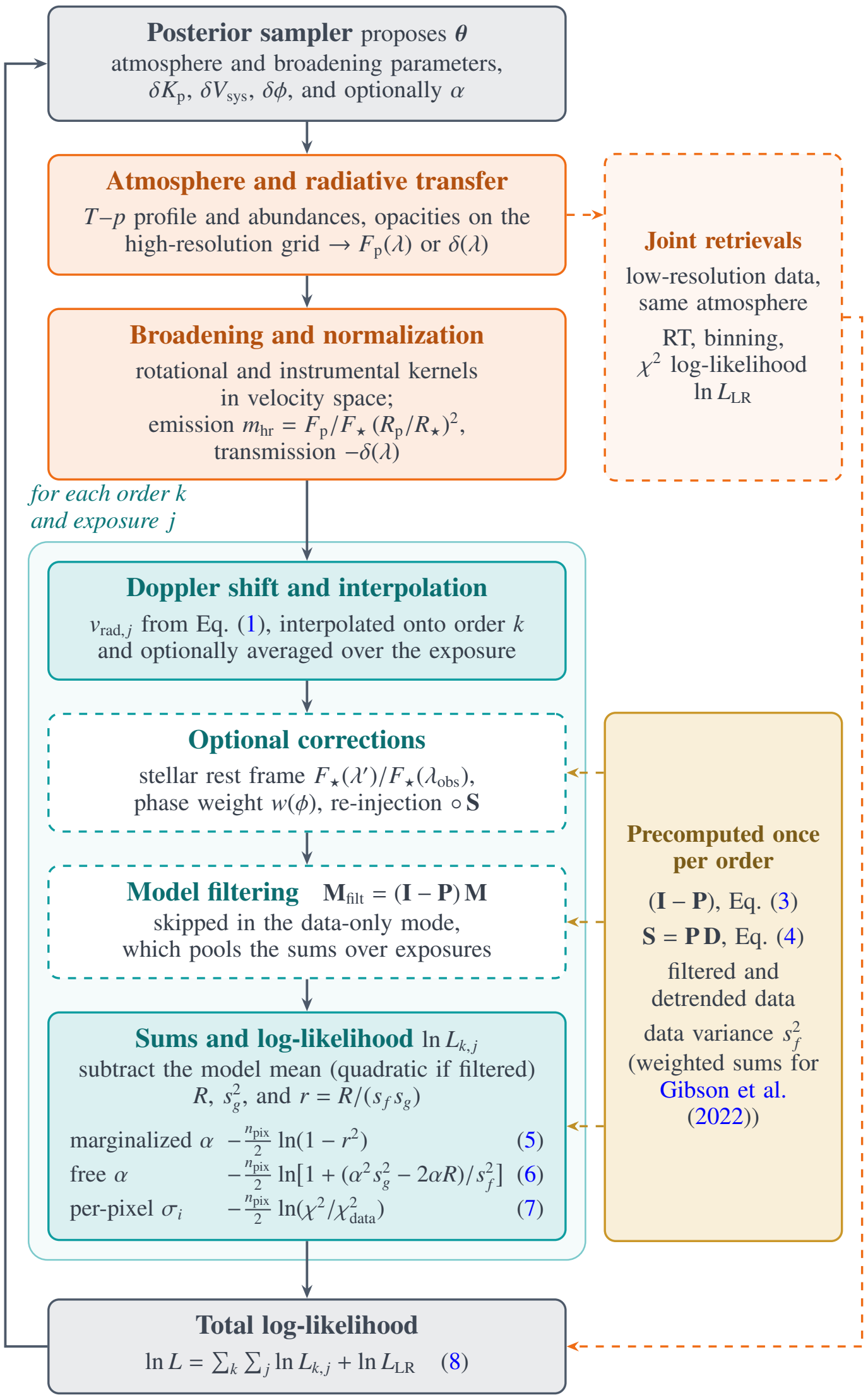
\begin{figure}[tbp]
\centering
\definecolor{bearorange}{HTML}{EF6C15}
\definecolor{bearteal}{HTML}{0F9DA0}
\definecolor{bearslate}{HTML}{4D5969}
\definecolor{beargold}{HTML}{E0A82E}
\resizebox{\columnwidth}{!}{%
\thickmuskip=5mu\relax\medmuskip=4mu\relax
\begin{tikzpicture}[
  node distance = 0.42cm and 0.5cm,
  base/.style       = {rectangle, rounded corners=4pt, line width=0.8pt,
                       inner xsep=8pt, inner ysep=6pt, font=\small,
                       align=center, text=bearslate!70!black},
  samplerbox/.style = {base, draw=bearslate, fill=bearslate!12, text width=6.2cm},
  fwdbox/.style     = {base, draw=bearorange, fill=bearorange!13, text width=6.2cm},
  llbox/.style      = {base, draw=bearteal, fill=bearteal!15, text width=6.2cm},
  optbox/.style     = {base, draw=bearteal, fill=white, dashed, text width=6.2cm},
  sidebox/.style    = {rectangle, rounded corners=4pt, line width=0.8pt, inner sep=0pt},
  sidetext/.style   = {font=\footnotesize, align=center, text width=2.55cm,
                       text=bearslate!70!black},
  arrow/.style  = {-{Stealth[length=5pt,width=4pt]}, line width=0.9pt, bearslate},
  oarrow/.style = {-{Stealth[length=4.5pt,width=3.6pt]}, line width=0.8pt,
                   dashed, bearorange},
  garrow/.style = {-{Stealth[length=4.5pt,width=3.6pt]}, line width=0.8pt,
                   dashed, beargold!85!black},
]
\newcommand{\ttl}[2]{\textcolor{#1}{\normalsize\bfseries #2}}
\newcommand{\sttl}[2]{\textcolor{#1}{\small\bfseries #2}}
\node[samplerbox] (sampler) {\ttl{bearslate!80!black}{Posterior sampler} proposes $\boldsymbol\theta$\\[1pt]
  atmosphere and broadening parameters,\\
  $\delta\Kp$, $\delta\Vsys$, $\delta\phi$, and optionally $\alpha$};
\node[fwdbox, below=of sampler] (rt) {\ttl{bearorange!75!black}{Atmosphere and radiative transfer}\\[3pt]
  $T$--$p$ profile and abundances, opacities on the\\
  high-resolution grid $\to F_\mathrm{p}(\lambda)$ or $\delta(\lambda)$};
\node[fwdbox, below=of rt] (broad) {\ttl{bearorange!75!black}{Broadening and normalization}\\[3pt]
  rotational and instrumental kernels\\
  in velocity space;\\
  emission $m_\mathrm{hr}=F_\mathrm{p}/F_\star\,(R_\mathrm{p}/R_\star)^2$,\\
  transmission $-\delta(\lambda)$};
\node[llbox, below=1.05cm of broad] (doppler) {\ttl{bearteal!70!black}{Doppler shift and interpolation}\\[3pt]
  $v_{\mathrm{rad},j}$ from Eq.~(\ref{eq:vrad}), interpolated onto order $k$\\
  and optionally averaged over the exposure};
\node[optbox, below=of doppler] (corr) {\ttl{bearteal!70!black}{Optional corrections}\\[3pt]
  stellar rest frame $F_\star(\lambda')/F_\star(\lambda_\mathrm{obs})$,\\
  phase weight $w(\phi)$, re-injection $\circ\,\mathbf{S}$};
\node[optbox, below=of corr] (filter) {\ttl{bearteal!70!black}{Model filtering}\quad
  $\mathbf{M}_\mathrm{filt}=\IP\,\mathbf{M}$\\[1pt]
  skipped in the data-only mode,\\
  which pools the sums over exposures};
\node[llbox, below=of filter] (ll) {\ttl{bearteal!70!black}{Sums and log-likelihood} $\ln L_{k,j}$\\[1pt]
  subtract the model mean (quadratic if filtered)\\
  $R$, $s_g^2$, and $r=R/(s_f s_g)$\\[4pt]
    \resizebox{\linewidth}{!}{\footnotesize\begin{tabular}{@{}l@{\;\;}l@{\;\;}r@{}}
  marginalized $\alpha$ & $-\tfrac{n_\mathrm{pix}}{2}\ln(1-r^2)$ & (\ref{eq:bear_loglike})\\[2pt]
  free $\alpha$ & $-\tfrac{n_\mathrm{pix}}{2}\ln\bigl[1+(\alpha^2 s_g^2-2\alpha R)/s_f^2\bigr]$ & (\ref{eq:free_alpha})\\[2pt]
  per-pixel $\sigma_i$ & $-\tfrac{n_\mathrm{pix}}{2}\ln(\chi^2/\chi^2_\mathrm{data})$ & (\ref{eq:gibson_bear})
  \end{tabular}}};
\node[samplerbox, below=0.75cm of ll] (total) {\ttl{bearslate!80!black}{Total log-likelihood}\\[3pt]
  $\ln L=\sum_{k}\sum_{j}\ln L_{k,j}+\ln L_\mathrm{LR}$\quad(\ref{eq:joint})};
\begin{scope}[on background layer]
  \node[draw=bearteal!55, fill=bearteal!4, line width=0.6pt, rounded corners=7pt,
        inner sep=7pt, fit=(doppler)(corr)(filter)(ll)] (loopbox) {};
\end{scope}
\node[anchor=south west, font=\footnotesize\itshape, text=bearteal!75!black, inner sep=1pt, align=left]
  at ([yshift=2pt]loopbox.north west) {for each order $k$\\and exposure $j$};
\node[sidebox, draw=bearorange, fill=bearorange!6, dashed,
      fit={([xshift=0.5cm]rt.north east) ([xshift=3.6cm]broad.south east)}] (lr) {};
\node[sidetext] at (lr.center) {\sttl{bearorange!75!black}{Joint retrievals}\\[3pt]
  low-resolution data,\\ same atmosphere\\[3pt] RT, binning,\\ $\chi^2$ log-likelihood\\ $\ln L_\mathrm{LR}$};
\node[sidebox, draw=beargold!90!black, fill=beargold!18,
      fit={([xshift=0.5cm]corr.north east) ([xshift=3.6cm]ll.south east)}] (pre) {};
\node[sidetext] at (pre.center) {\sttl{beargold!55!black}{Precomputed once\\per order}\\[5pt]
  $\IP$, Eq.~(\ref{eq:projection})\\[3pt]
  $\mathbf{S}=\mathbf{P}\,\mathbf{D}$, Eq.~(\ref{eq:reinject_S})\\[3pt]
  filtered and\\ detrended data\\[3pt]
  data variance $s_f^2$\\ (weighted sums for\\ \citet{Gibson2022})};
\draw[arrow] (sampler) -- (rt);
\draw[arrow] (rt) -- (broad);
\draw[arrow] (broad) -- (doppler);
\draw[arrow] (doppler) -- (corr);
\draw[arrow] (corr) -- (filter);
\draw[arrow] (filter) -- (ll);
\draw[arrow] (ll) -- (total);
\draw[arrow] (total.west) -- ++(-0.55,0) |- (sampler.west);
\draw[oarrow] (rt.east) -- (rt.east -| lr.west);
\draw[oarrow] (lr.east) -- ++(0.25,0) |- (total.east);
\draw[garrow] (corr.east -| pre.west) -- (corr.east);
\draw[garrow] (filter.east -| pre.west) -- (filter.east);
\draw[garrow] (ll.east -| pre.west) -- (ll.east);
\end{tikzpicture}}
\caption{Workflow of one high-resolution likelihood evaluation in \bear. Orange boxes are computed once per trial atmosphere and green boxes inside the per-order, per-exposure loop (shaded region); dashed borders mark optional steps. The quantities in the yellow box are computed once per order at initialization, and the low-resolution branch enters only joint retrievals (Sect.~\ref{sec:joint_like}). The configuration selects one of the three likelihood modes (Sect.~\ref{sec:like}).}
\label{fig:workflow}
\end{figure}

This section describes the high-resolution extension of \bear for emission and transmission spectra and notes where it differs from \citet{Brogi2019}, \citet{Gibson2022}, and the \chimera implementation of \citet{Line2021}. Figure~\ref{fig:workflow} shows one likelihood evaluation. Apart from the atmospheric structure, all of its steps run on the GPU, and an equivalent OpenMP version runs on the CPU (Sect.~\ref{sec:cuda}).

\subsection{High-resolution forward model}
\label{sec:hr_forward}

Each high-resolution observation (i.e., one time series of spectra, such as a single night or the pre-eclipse sequence of an emission observation) is read from its own observation file. When a retrieval contains at least one such observation, \bear computes the model on a separate high-resolution spectral grid, a set of wavenumber points with a nominal resolving power set by the user (typically $R\gtrsim 2.5\times10^5$). This grid covers only the wavelength range of the high-resolution data, extended by 300\,km\,s$^{-1}$ on both sides so that pixels Doppler-shifted by up to this velocity never fall outside the model. Low-resolution observations in the same retrieval use the usual low-resolution grid, which covers their own wavelength range at a lower resolving power, so that a joint retrieval computes two spectra of the same atmosphere instead of one high-resolution spectrum over both wavelength ranges. Both grids are subsets of the wavenumber points on which the opacities are tabulated at a constant step of 0.01\,cm$^{-1}$, so the spacing of the high-resolution grid only approximates the nominal $R$. Dayside emission is modeled with the \texttt{phase\_curve} forward model and transits with the transmission model, which use the same temperature, chemistry, gas-opacity, and radiative-transfer routines as the low-resolution models (Sect.~\ref{sec:bearold}). Cloud opacity and the \texttt{stellar\_contamination} module, which models the transit light-source effect of unocculted spots and faculae \citep{Rackham2018}, currently act on the low-resolution spectrum only. The high-resolution spectrum can use its own set of opacity tables, for example to omit a species without lines in the observed range. 
 
The emission model returns the planetary flux $F_\mathrm{p}(\lambda)$, which is broadened (Sect.~\ref{sec:broadening}) and then divided by a stellar spectrum computed on the same grid to form the dimensionless planet-to-star contrast $m_\mathrm{hr}=F_\mathrm{p}/F_\star\,(R_\mathrm{p}/R_\star)^2$. The transmission model returns the transit depth $\delta(\lambda)=[R_\mathrm{p}(\lambda)/R_\star]^2$. Any temperature profile of Sect.~\ref{sec:bear_temperature} can be used. Three-dimensional effects on high-resolution spectra in transmission \citep[e.g.,][]{Wardenier2023} and emission \citep[e.g.,][]{Wardenier2025_3d, Wardenier2026_windfields} are beyond the scope of this paper but are planned for future versions of \bear.

\subsection{Doppler shifting and data framing}
\label{sec:doppler}

Assuming a circular orbit, the planet's radial velocity at orbital phase $\phi_j$ is
\begin{equation}
  v_{\mathrm{rad},j} = \bigl(\Kp^\mathrm{ref}+\delta\Kp\bigr)
    \sin\,\bigl[2\pi(\phi_j + \delta\phi)\bigr]
    + \Vsys^\mathrm{ref}+\delta\Vsys + v_{\mathrm{bary},j}\,,
  \label{eq:vrad}
\end{equation}
where $\Kp^\mathrm{ref}$ and $\Vsys^\mathrm{ref}$ are optional reference values stored in the observation file, $\delta\Kp$ and $\delta\Vsys$ are the retrieved offsets, $\delta\phi$ is an optional orbital-phase offset, and $v_{\mathrm{bary},j}$ is the per-exposure barycentric velocity. Sampling offsets around literature values, as in \chimera \citep{Line2021} and \poseidon \citep{Wang2025}, keeps the sampler near the signal peak; if no reference values are given, $\Kp$ and $\Vsys$ are retrieved directly. The optional phase offset absorbs a shift of the radial-velocity curve in phase, which can come from an error in the adopted mid-transit time, for example after an ephemeris is propagated over many orbits, or from a small orbital eccentricity. For planets with precise ephemerides, such as WASP-77A\,b, it can be fixed at zero (Appendix~\ref{app:13co}). Barycentric velocities are supplied in the observation file or set to zero if the data reduction has already removed them. Many data reduction pipelines shift the wavelength solution to the stellar rest frame \citep[e.g.,][]{Hoeijmakers2024}; in that case $\Vsys^\mathrm{ref}$ must be set to zero, and the prior on $\delta\Vsys$, which then measures any remaining offset, is centered on zero, as in \citet{Wang2025}.

The model is evaluated at the shifted wavelengths $\lambda' = \lambda_\mathrm{obs}/(1 + v_\mathrm{rad}/c)$ by linear interpolation onto the observed grid of each order. Only the planetary flux should be shifted, but \bear shifts the emission contrast $m_\mathrm{hr}$, which also contains the stellar spectrum. It therefore multiplies the shifted contrast by $F_\star(\lambda')/F_\star(\lambda_\mathrm{obs})$, which restores the stellar spectrum to the observed wavelengths. \chimera instead divides by a stellar spectrum smoothed with a Gaussian of $\sigma=200$ model pixels, which removes the stellar lines \citep[][public example code]{Line2021}. \bear offers the same option, with three passes of a box filter that approximate a Gaussian of a chosen width in model pixels.

\subsection{Velocity broadening}
\label{sec:broadening}

Before cross-correlation, the model is broadened to the observed line width, which combines the planet's rotation with the instrumental line-spread function \citep{Brogi2016,Spring2022}. The broadening width can either be fixed from known values, as in \citet{Line2021}, or retrieved as a free parameter, as in \citet{Gibson2022}. \bear allows both choices and provides two broadening modules, which act only on the high-resolution spectrum and convolve it in velocity space. Because the high-resolution grid is a subset of the wavenumber points of the opacity tables (Sect.~\ref{sec:hr_forward}), its velocity step is not constant; for the IGRINS range at $R=2.5\times10^5$ it varies between about 0.6 and 1.2\,km\,s$^{-1}$. Each kernel is therefore evaluated at the actual velocity offsets of the neighboring grid points and normalized to unit area at every point, so that the broadening has the same width in km\,s$^{-1}$ at all wavelengths. Converting the kernel width into grid points with a single representative step would instead change the applied width by up to a factor of two across the IGRINS band. If neither module is selected, the model is not broadened.

The first module, \texttt{velocity\_broadening}, applies the analytic rotational kernel of \citet[][Eq.~18.14]{Gray2005},
\begin{equation}
  G(x) = \frac{2(1-\epsilon)\sqrt{1-x^2}
    + \tfrac{\pi}{2}\,\epsilon\,(1-x^2)}
    {\pi\,v\sin i\,(1-\epsilon/3)}\,,
  \qquad |x|\le1\,,
  \label{eq:rot_kernel}
\end{equation}
with $x = \Delta v/(v\sin i)$ and $\epsilon$ a linear limb-darkening coefficient, followed by a Gaussian of standard deviation $\sigma_\mathrm{tot}=\sqrt{\sigma_\mathrm{inst}^2 + v_\mathrm{wind}^2}$, where $\sigma_\mathrm{inst}$ is the standard deviation of the instrumental profile and $v_\mathrm{wind}$ captures any additional broadening by winds or turbulence. Because $\epsilon$ and $v\sin i$ are partly degenerate, $\epsilon$ is usually fixed and $v\sin i$ retrieved. The scheme matches that of \citet{Line2021}, who set $\epsilon=0$, as we do in Sect.~\ref{sec:w77_setup}.

The second available module is intended for transmission spectroscopy at high resolution \citep{Brogi2016} and will be described in a forthcoming publication.

The finite exposure time adds a broadening that differs between exposures because the planet's velocity changes during each exposure, and the recorded spectrum is the time average of the moving line profile. When the orbital period $P$ and the exposure times $t_\mathrm{exp}$ are given in the observation file, \bear replaces the point interpolation of Sect.~\ref{sec:doppler} with a top-hat average \citep[cf.][]{Pelletier2025} centered on $v_{\mathrm{rad},j}$, of full width $\Delta v = (\Kp^\mathrm{ref}+\delta\Kp)\,|\cos[2\pi(\phi_j+\delta\phi)]|\,(2\pi/P)\,t_\mathrm{exp}$, which is largest near conjunction and vanishes at quadrature.

\subsection{Fast model filtering and re-injection}
\label{sec:filtering}

Data detrending with \sysrem \citep{Tamuz2005} or PCA \citep{deKok2013} removes the dominant temporal systematics together with part of the planet signal, so that every trial model must be distorted in the same way \citep{Brogi2019,Gibson2020}. Instead of re-running the detrending for every model \citep{Brogi2019,Line2021}, we adopt the fast filtering of \citet{Gibson2022}. For one order, we write the data as a matrix $\mathbf{D}$ with one row per exposure and one column per pixel ($n_\mathrm{exp}\times n_\mathrm{pix}$ for $n_\mathrm{exp}$ exposures and $n_\mathrm{pix}$ pixels), and the $n_\mathrm{comp}$ \sysrem or PCA basis vectors, augmented with a column of ones that also removes the per-pixel temporal mean, as the columns of $\mathbf{U}$. The detrended data are then $\IP\,\mathbf{D}$, with $\mathbf{I}$ the $n_\mathrm{exp}\times n_\mathrm{exp}$ identity and the projection matrix
\begin{equation}
\mathbf{P}=\mathbf{U}\bigl(\mathbf{U}^\top\boldsymbol{\Lambda}^2\mathbf{U}\bigr)^{-1}
    \mathbf{U}^\top\boldsymbol{\Lambda}^2\,,
  \label{eq:projection}
\end{equation}
where $\boldsymbol{\Lambda}=\mathrm{diag}(1/\bar\sigma_j)$ optionally down-weights each exposure $j$ by its mean uncertainty $\bar\sigma_j$ \citep[][Eq.~7]{Gibson2022}. Applying the same operator to the model matrix $\mathbf{M}$ ($n_\mathrm{exp}\times n_\mathrm{pix}$), which holds the Doppler-shifted model of every exposure of the order (Sect.~\ref{sec:doppler}), distorts the model as the detrending distorts the planet signal, $\mathbf{M}_\mathrm{filt}=\IP\,\mathbf{M}$. Because $\IP$ depends only on $\mathbf{U}$ and $\boldsymbol{\Lambda}$, \bear computes it once per order at initialization, and each likelihood evaluation costs one $(n_\mathrm{exp}\times n_\mathrm{exp})$ matrix product per pixel column.

In a data-only mode (\texttt{\#filter\_model~0}), \bear applies $\IP$ to the data but not to the model, from which only the temporal mean of each pixel is removed. The filtered data and the model then both have a zero temporal mean at every pixel, so model components that are constant in time, such as a smooth continuum, cancel from the sum over all exposures, whereas the moving lines do not. \bear therefore accumulates the cross-correlation sums over all exposures of an order before taking the logarithm. Because the model is not distorted like the planet signal, this mode serves only as a diagnostic.

The cross-correlation likelihoods of Sect.~\ref{sec:like} compare the data with the model scaled by a factor $\alpha$, the line contrast. When $\alpha$ is eliminated analytically, the likelihood depends only on the shape of the cross-correlation and cannot distinguish models whose lines differ mainly in depth \citep{Brogi2019}. A free $\alpha$ recovers this information only if model and data share a common scale, so that $\alpha\approx1$ for the correct atmosphere. This holds for transmission, where \bear compares the relative in-transit flux change $-\delta(\lambda)$. For dayside emission, the contrast $m_\mathrm{hr}$ is relative to the stellar continuum, whereas the planet signal in the observed residuals is further modulated by the stellar, telluric, and instrumental throughput of each pixel and exposure. \citet{Line2021} addressed this in \chimera by re-injecting the model into the low-rank reconstruction of the data and re-running the PCA at every likelihood call. \citet{Wang2025} do the same in the PCA mode of \poseidon, and in its fast-filtering mode they multiply the low-rank reconstruction by the model and filter the product with the fixed basis.

\bear combines this re-injection with the fixed filter. At initialization, it stores the part of the data that the filter removes,
\begin{equation}
  \mathbf{S} = \mathbf{D} - \IP\,\mathbf{D} = \mathbf{P}\,\mathbf{D}\,,
  \label{eq:reinject_S}
\end{equation}
which corresponds to the low-rank reconstruction of \chimera. At each likelihood call, \bear Doppler-shifts the dimensionless model into $\mathbf{M}$, multiplies it element-wise by $\mathbf{S}$ to convert it to the units of the data, and filters it, $\mathbf{M}_\mathrm{filt}=\IP(\mathbf{M}\circ\mathbf{S})$. A model component that is constant in time at a given pixel only rescales the corresponding column of $\mathbf{S}$, which lies in the span of $\mathbf{U}$, so that it cancels because $\IP\,\mathbf{P}=0$, and only the moving lines survive. Model and data then share a common scale, and a free $\alpha$ with a bounded prior restores sensitivity to the absolute abundances. The scheme thus corresponds to \chimera with the basis held fixed, as in the fast filtering of \citet{Gibson2022}. Re-injection (\texttt{\#reinject\_model~1}) requires model filtering and is informative only with the free-$\alpha$ likelihoods of Eqs.~(\ref{eq:free_alpha}) and (\ref{eq:gibson_bear}).
 
For dayside emission, the visible fraction of the hot dayside changes with orbital phase. \bear can weight the model of each exposure before filtering by the phase function $w(\phi)=\{\tfrac12[1+\cos(2\pi(\phi+\delta\phi)-\pi)]\}^2$ \citep{Herman2022,Hoeijmakers2024,Pelletier2025}, which is unity at secondary eclipse, 1/4 at quadrature, and zero at transit. Without it, every exposure carries equal weight, including those far from eclipse that contain little planetary signal.

\begin{figure*}[ht]
  \centering
  \includegraphics[width=\columnwidth]{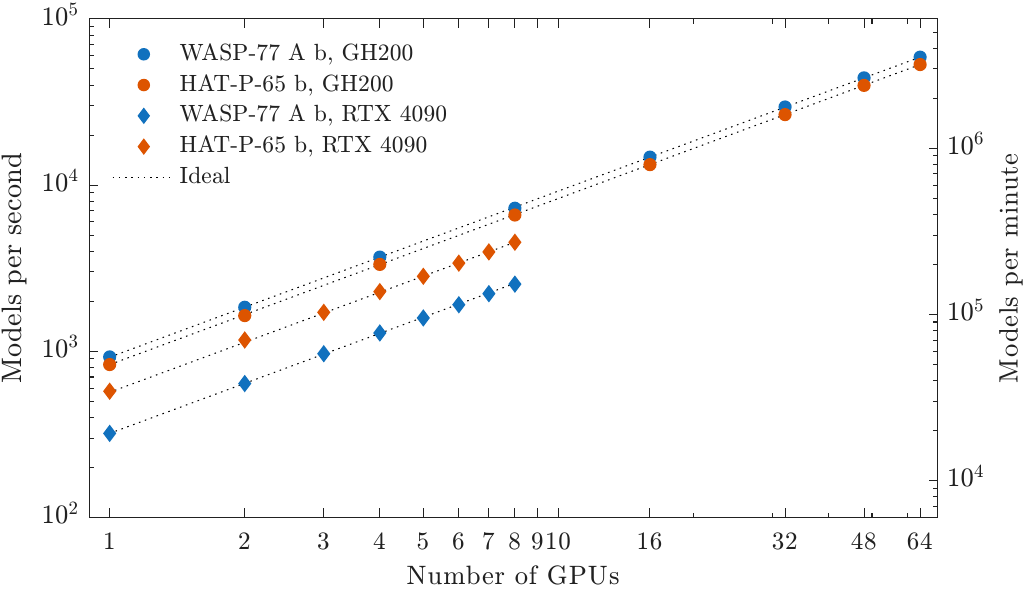}
  \includegraphics[width=0.9\columnwidth]{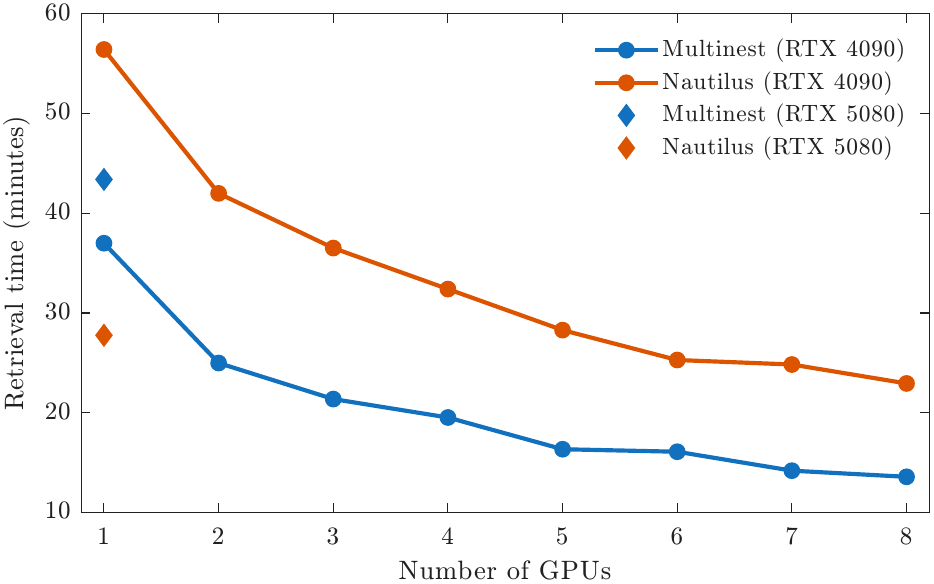}
  \vspace{-0.2cm}
  \caption{Computational performance of \bear on multiple GPUs. \textit{Left:} Scaling of the forward model throughput with the number of GPUs. Blue and orange refer to the high-resolution forward model for WASP-77A\,b and the JWST NIRSpec/PRISM model for HAT-P-65\,b, respectively. Circles mark the throughput on Daint (CSCS) with NVIDIA GH200 GPUs, while diamonds show the results for NVIDIA RTX 4090 GPUs on UBELIX. The dotted lines show the ideal, linear scaling. \textit{Right:} Scaling of the retrieval time with the number of RTX 4090 GPUs on UBELIX for the WASP-77A\,b IGRINS retrieval. Blue marks \multinest and orange \nautilus. The diamonds indicate the results for a desktop computer with a single NVIDIA GeForce RTX 5080 GPU.}
  \label{fig:gpu_scaling}
\end{figure*}

\subsection{Cross-correlation log-likelihood}
\label{sec:like}

\bear provides three cross-correlation log-likelihoods 
(see Fig.~\ref{fig:workflow}).
They follow from a Gaussian noise model with a model scale factor and a global noise scale and differ only in which of these nuisance parameters are eliminated analytically \citep{Brogi2019,Gibson2020,Gibson2022}. We therefore give only the forms that \bear evaluates and the choices specific to \bear. In each order and exposure, $f_i$ and $g_i$ denote the detrended data and the Doppler-shifted, broadened, and filtered model at pixel $i$, $n_\mathrm{pix}$ is the number of pixels, and $\bar f_i$ and $\bar g_i$ are $f_i$ and $g_i$ with their mean subtracted. From these, \bear forms $s_f^2=\tfrac{1}{n_\mathrm{pix}}\sum_i\bar f_i^2$, $s_g^2=\tfrac{1}{n_\mathrm{pix}}\sum_i\bar g_i^2$, and $R=\tfrac{1}{n_\mathrm{pix}}\sum_i\bar f_i\bar g_i$, and the correlation coefficient $r=R/(s_f s_g)$. Residual continuum curvature, such as the blackbody slope shifted by the planet's velocity, survives the temporal filter and could dominate $s_g^2$ without contributing to $R$. Therefore, when the model is filtered (Sect.~\ref{sec:filtering}), \bear subtracts a second-order polynomial in pixel index instead of the mean, from the data once at initialization and from the model at every call. 

Without a prior for $\alpha$, \bear uses the mapping of \citet{Zucker2003},
\begin{equation}
  \ln\mathcal{L}_{k,j} = -\frac{n_\mathrm{pix}}{2}\ln\bigl(1-r^2\bigr)\,,
  \qquad
  \ln\mathcal{L}=\sum_{k}\sum_{j}\ln\mathcal{L}_{k,j}\,,
  \label{eq:bear_loglike}
\end{equation}
which is the likelihood of \citet[][Eq.~9]{Brogi2019} maximized over $\alpha$. Because each order and exposure contributes its own $\ln(1 - r^2)$ term, every term is implicitly fitted with an independent scale factor of either sign. This is inherent to the mapping and shared by other implementations. The retrieval is then insensitive to the absolute line contrast of the model, which in emission is degenerate with the temperature gradient and with dayside inhomogeneity, and in transmission with the scale height. 
 
When a prior for $\alpha$ is present, which is the common case and the one used for all retrievals in this paper, \bear instead samples a single $\alpha$ shared by all orders, exposures, and high-resolution observations, as in \citet{Gibson2020}, \citet{Pelletier2021}, and \citet{Line2021},
\begin{equation}
  \ln\mathcal{L}_{k,j} = -\frac{n_\mathrm{pix}}{2}\ln\Bigl[1 + \frac{\alpha^2 s_g^2 - 2\alpha R}{s_f^2}\Bigr]\,,
  \label{eq:free_alpha}
\end{equation}
which is the likelihood of \citet[][Eq.~9]{Brogi2019} with a sampled scale factor. This mode preserves the absolute line contrast, is required by the emission re-injection scheme (Sect.~\ref{sec:filtering}), and suits joint retrievals, in which the low-resolution spectrum fixes the temperature gradient and with it $\alpha$ (Sect.~\ref{sec:w77_joint}). A value of $\alpha$ different from unity then shows how far the line contrast of the one-dimensional model departs from that of the data, for example because of an inhomogeneous dayside. When the observation file also provides per-pixel uncertainties $\sigma_i$, \bear uses the $\beta$-marginalized likelihood of \citet[][Eq.~4]{Gibson2022},
\begin{equation}
  \ln\mathcal{L}_{k,j} = -\frac{n_\mathrm{pix}}{2}\ln\frac{\chi^2(\alpha)}{\chi^2_\mathrm{data}}\,,
  \label{eq:gibson_bear}
\end{equation}
where $\chi^2(\alpha)$ is the inverse-variance-weighted $\chi^2$ of the mean-subtracted residuals and $\chi^2_\mathrm{data}=\chi^2(0)$. Its data-only sums are precomputed, and the filtered residuals enter with the original $\sigma_i$, which assume Gaussian noise, a good approximation for photon-limited data at high count rates.

\subsection{Joint high- and low-resolution likelihood}
\label{sec:joint_like}

When a retrieval contains both high-resolution and low-resolution observations, each data set is evaluated with its own likelihood and the log-likelihoods are summed \citep{Brogi2017,Brogi2019,Smith2024},
\begin{equation}
  \ln\mathcal{L}_\mathrm{joint} = \ln\mathcal{L}_\mathrm{HR} + \ln\mathcal{L}_\mathrm{LR}\,,
  \label{eq:joint}
\end{equation}
where $\ln\mathcal{L}_\mathrm{HR}$ is the cross-correlation log-likelihood of Sect.~\ref{sec:like} summed over all orders, exposures, and high-resolution observations. $\ln\mathcal{L}_\mathrm{LR}$ is the Gaussian log-likelihood of the low-resolution data, including the normalization term $-\tfrac12\sum_i\ln(2\pi\sigma_i^2)$, which matters when the errors are inflated as in \citet{Line2015} and \citet{Kitzmann2020}. Both terms are computed from the same trial atmosphere on their respective spectral grids, so the abundances and the temperature profile are shared. The velocities, broadening parameters, and $\alpha$ enter only the high-resolution term and are shared by all high-resolution observations, while each observation file carries its own reference velocities.

\section{GPU implementation and computational cost}
\label{sec:cuda}

The design goal of \bear is to complete nested-sampling retrievals, which require $10^5$ to $10^6$ forward models, in minutes to hours on a single desktop computer. A low-resolution forward model with 70 layers and $\sim7000$ wavelength points took about 20\,ms on a consumer GPU in \citet{Kitzmann2020}. Since the original release of \heliosr, the computational performance of the model has been steadily improved to a point where retrieval calculations that took about two hours in \citet{Kitzmann2020} can now be completed within roughly two minutes.

At high resolution, a line-by-line model spectrum at $R\gtrsim2.5\times10^5$ contains $10^5$ to $10^6$ points per instrument band, and a data set of $\sim10^2$ exposures over several tens of orders compares $\sim10^7$ pixels per likelihood call. The new CUDA kernels for the high-resolution data processing have been highly optimized to make effective use of the GPU architecture. Data transfers between the main memory and the GPU are comparatively slow. To limit them, only the atmospheric structure, that is, the temperature profile, the chemistry, and the hydrostatic structure, is computed on the CPU, while everything else, including the calculation of the model likelihood itself, is done on the GPU. The only value returned to the CPU at the end is the likelihood of the model for the given parameter set, which minimizes the data exchange between CPU and GPU and enhances the computational performance.

Furthermore, with the parallel build of \multinest that uses the message passing interface (MPI), each MPI process holds its own instance of the \bear forward model on its own GPU. This allows several parameter sets to be evaluated simultaneously, so that the forward model throughput scales linearly with the number of available GPUs. This scaling was tested on UBELIX, the high-performance computing cluster of the University of Bern, which is equipped with consumer-grade NVIDIA RTX 4090 GPUs, and on Daint at the Swiss National Supercomputing Centre (CSCS), which contains server-grade NVIDIA GH200 GPUs. We tested two cases, the high-resolution emission case of WASP-77A\,b and a JWST NIRSpec/PRISM observation of HAT-P-65\,b (Prinoth et al., in prep.). The latter was evaluated at a resolving power of 20\,000 in the forward model.

Figure~\ref{fig:gpu_scaling} shows the resulting number of forward model evaluations per second as a function of the number of GPUs. On UBELIX, the number of RTX 4090 GPUs that can be used simultaneously is limited to eight.

As Fig.~\ref{fig:gpu_scaling} indicates, the scaling on both clusters follows the ideal case, with the number of forward model evaluations increasing linearly with the number of GPUs. The scaling is linear because each GPU computes its own forward model and the GPUs do not need to communicate with each other. On Daint, both scenarios show roughly the same throughput. On a single GPU, one call requires a little more than 1\,ms, providing close to 1000 forward models per second, or about 60\,000 per minute. With 64 GPUs, \bear generates more than three million forward models per minute in both scenarios. If the sampler kept pace, this throughput would allow full retrievals to be performed in less than one minute and would also allow machine-learning emulators to be trained on the fly, without the need to store an extensive training data set.

The consumer-grade GPUs on UBELIX are slower, as expected, by a factor of about three for the high-resolution case and about 1.5 for the HAT-P-65\,b case. Nonetheless, with the eight available GPUs on UBELIX, between 150\,000 and 270\,000 models can be generated per minute, which again shows that retrievals can be done in a matter of minutes rather than hours.

Table~\ref{tab:timing} lists the cost of a full WASP-77A\,b retrieval of Sect.~\ref{sec:validation}. One likelihood evaluation, forward model included, costs 3.9\,ms on the GPU and 680\,ms on an AMD Ryzen 9 CPU of the same desktop computer, so that the 573\,014 evaluations of the full 17-parameter retrieval took 42 minutes instead of about 4.5\,days. For comparison, the \chimera analysis of the same data with 500 live points took about 3 days for about 270\,000 likelihood evaluations on 24 CPUs and one GPU \citep{Line2021}, and a typical \poseidon emission retrieval with 12 parameters and 800 live points about 7\,h on 24 Intel Xeon E5-2670 CPUs \citep{Wang2025}. Since runtimes depend on the number of live points, parameters, exposures, and orders, these comparisons are indicative only. 

At this cost, the benchmark retrievals with \multinest and \nautilus and the five reduced-model retrievals of Appendix~\ref{app:13co} took 2.6\,h together, and each injection--recovery retrieval of Appendix~\ref{app:injection} took 0.4 to 1.5\,h.

\begin{table}[ht]
\centering
\caption{Computational cost of the WASP-77A\,b IGRINS retrieval (Sect.~\ref{sec:validation}) on a single NVIDIA GeForce RTX 5080 GPU.}
\label{tab:timing}
\small
\begin{tabular}{lc}
\toprule
Quantity & Value \\
\midrule
High-resolution grid points                      & 186\,552 \\
Likelihood evaluation incl.\ forward model, GPU  & 3.9\,ms \\
Likelihood evaluation, CPU (16 cores)          & 680\,ms \\
Live points                                      & 800 \\
Likelihood evaluations, \multinest               & 573\,014 \\
Likelihood evaluations, \nautilus                & 176\,100 \\
Wall time, \multinest                            & 0.70\,h \\
Wall time, \nautilus                             & 0.46\,h \\
Wall time, joint IGRINS$+$NIRSpec, \multinest    & 0.94\,h \\
\midrule
\citet{Line2021}, 24 CPUs $+$ 1 GPU & $\sim$3\,d \\
\citet{Wang2025}, 24 CPUs & $\sim$7\,h \\
\bottomrule
\end{tabular}
\tablefoot{The likelihood cost is the median over the parameter vectors visited by \multinest in a short run with the benchmark configuration and 50 live points. The joint retrieval uses 1000 live points (Sect.~\ref{sec:w77_joint}).}
\end{table}

We also measured the speed-up of a full retrieval with up to eight GPUs on UBELIX. While the forward model evaluations scale linearly with the number of GPUs, Fig.~\ref{fig:gpu_scaling} shows that the full retrieval does not, with either \multinest or \nautilus.
 
The speed-up diminishes quickly with an increasing number of GPUs, and doubling the number of GPUs from four to eight shortens the retrieval time by only 30\%; with all eight GPUs, a \multinest retrieval takes about 14 minutes. On UBELIX, unlike on the desktop (Table~\ref{tab:timing}), \nautilus is slower than \multinest. The training of its neural networks runs on the CPU, and the CPUs of the UBELIX GPU nodes are considerably slower than that of the desktop. The retrieval is thus limited by the samplers themselves, which cannot supply parameter combinations fast enough to use all available GPUs. Exploiting the full speed of \bear will likely require samplers that propose many parameter sets in parallel, such as GPU-based samplers.

\section{Validation on WASP-77A\,b}
\label{sec:validation}

In the following we validate the framework on the dayside emission of the hot Jupiter WASP-77A\,b ($R_\mathrm{p}=1.21\,R_\mathrm{J}$, $M_\mathrm{p}=1.76\,M_\mathrm{J}$, and $P=1.36$\,d, \citealt{Maxted2013}; $T_\mathrm{eq}\approx1740$\,K, \citealt{Line2021}), which has been observed at high resolution with IGRINS \citep{Park2014,Mace2018} on Gemini South \citep{Line2021} and at low resolution with JWST NIRSpec/G395H \citep{August2023}. \citet{Smith2024} retrieved both data sets separately and jointly with a common setup, and in all cases they find a sub-solar metallicity and a near-solar C/O ratio. The IGRINS data have also served as a benchmark for \poseidon \citep{Wang2025}. We repeated the three retrievals of \citet{Smith2024} that use the pre-eclipse IGRINS sequence with their parameterization, species, and priors, so that differences in the results can be attributed to the codes. Figure~\ref{fig:w77_summary} and Table~\ref{tab:w77} summarize the results, Appendices~\ref{app:robust} and \ref{app:injection} test the IGRINS retrieval, and Appendix~\ref{app:comparison} compares every retrieved parameter with \citet{Smith2024}. All retrievals used \multinest unless stated otherwise.

\subsection{Atmosphere model and priors}
\label{sec:w77_setup}

As in \citet{Smith2024}, we described the thermal structure with the six-parameter profile of \citet{Madhusudhan2009} and the composition with constant-with-altitude volume mixing ratios over an H$_2$/He background with $n_\mathrm{He}/n_\mathrm{H_2}=0.170$. In this profile, the temperature rises from $T_0$ at the top of the model atmosphere, $P_0=10^{-6}$\,bar, as $T=T_0+[\ln(P/P_0)/\alpha_1]^2$ down to $P_1$ and as $T=T_2+[\ln(P/P_2)/\alpha_2]^2$ down to $P_3$, below which it is isothermal, with $T_2$ set by continuity at $P_1$. The slope coefficients $\alpha_1$ and $\alpha_2$ (in K$^{-1/2}$) set the temperature gradients of the two upper layers, with larger values giving shallower gradients. The IGRINS and joint retrievals include \ce{H2O}, CO, \ce{CH4}, \ce{H2S}, \ce{NH3}, HCN, and \ce{CO2}, whereas the NIRSpec retrieval includes only \ce{H2O}, CO, and \ce{CO2}. The $^{13}$CO isotopologue is parameterized as the ratio offset $[\mathrm{^{13}CO/^{12}CO}] = \log_{10}(\mathrm{^{13}CO/^{12}CO})_\mathrm{p} - \log_{10}(\mathrm{^{13}CO/^{12}CO})_\oplus$ with $(\mathrm{^{13}CO/^{12}CO})_\oplus = 1/89$, which is also the solar value \citep{Meibom2007}, so that the $^{13}$CO abundance follows from the CO abundance and the retrieved ratio instead of being fitted as an independent mixing ratio. We used the \ce{H2O} POKAZATEL \citep{Polyansky2018}, CO and $^{13}$CO \citep{Li2015}, \ce{CH4} YT34to10 \citep{Yurchenko2017}, \ce{H2S} AYT2 \citep{Azzam2016}, \ce{NH3} CoYuTe \citep{Coles2019}, HCN \citep{Harris2006}, and \ce{CO2} UCL-4000 \citep{Yurchenko2020} line lists, together with H$_2$--H$_2$, H$_2$--He, and H--He collision-induced absorption \citep{Karman2019}, all in \heliosk format. The atmosphere was divided into 100 levels between 400 and $10^{-6}$\,bar, equidistant in log-pressure, and the emission was computed with the short-characteristics solver of Sect.~\ref{sec:bear_forward}. We fixed the surface gravity and the planet-to-star radius ratio at the values of \citet{Maxted2013}, $\log g=3.441$ and $R_\mathrm{p}/R_\star=0.1301$, and used a PHOENIX stellar spectrum \citep{Husser2013} with $T_\mathrm{eff}=5605$\,K \citep{Mortier2013} and $\log g=4.33$ \citep{Maxted2013}. The priors, listed in Table~\ref{tab:w77_priors}, are those of \citet{Smith2024}; they constrain the three pressures of the temperature profile only through their ranges, so that no ordering of $P_1$, $P_2$, and $P_3$ is imposed (see Sect.~\ref{sec:disc_codes}).

\begin{table}[ht]
\centering
\caption{Free parameters and priors of the WASP-77A\,b retrievals.}
\label{tab:w77_priors}
\footnotesize
\setlength{\tabcolsep}{3pt}
\begin{tabular}{p{4.0cm}lc}
\toprule
Parameter & Prior & Used in \\
\midrule
$X_\mathrm{H_2O}$, $X_\mathrm{CO}$, $X_\mathrm{CO_2}$ & log-U $[-12,0]$ & all \\[2pt]
$X_\mathrm{CH_4}$, $X_\mathrm{H_2S}$, $X_\mathrm{NH_3}$, $X_\mathrm{HCN}$ & log-U $[-12,0]$ & HR, joint \\[2pt]
$[\mathrm{^{13}CO/^{12}CO}]$ & U $[-5,5]$ & all \\[2pt]
$T_0$ (K) & U $[500,2500]$ & all \\[2pt]
$P_1$, $P_2$ (bar) & log-U $[-5.5,2.5]$ & all \\[2pt]
$P_3$ (bar) & log-U $[-2,2]$ & all \\[2pt]
$\alpha_1$, $\alpha_2$ (K$^{-1/2}$) & U $[0.02,2]$ & all \\[2pt]
$\delta\Kp$, $\delta\Vsys$ (km\,s$^{-1}$) & U $[-20,20]$ & HR, joint \\[2pt]
$\alpha$ & log-U $[0.01,100]$ & HR, joint \\
\bottomrule
\end{tabular}
\tablefoot{U denotes uniform and log-U log-uniform priors. $\alpha_1$ and $\alpha_2$ are parameters of the temperature profile. Abundances $X$ and pressures $P_i$ are sampled in $\log_{10}$, so that the tabulated ranges apply to $\log_{10}X$ and $\log_{10}P_i$. HR denotes the IGRINS retrieval and joint denotes the IGRINS$+$NIRSpec retrieval, both with 17 free parameters; the NIRSpec retrieval has 10. The fixed parameters are $v\sin i = 4.52$\,km\,s$^{-1}$, $\sigma_\mathrm{inst}=2.83$\,km\,s$^{-1}$, $\epsilon = 0$, $v_\mathrm{wind}=0$, $\delta\phi = 0$, and the NIRSpec detector offset (0). Velocities are offsets from $\Kp^\mathrm{ref}=192.06$ and $\Vsys^\mathrm{ref}=1.6845$\,km\,s$^{-1}$.}
\end{table}

\begin{table*}[ht]
\centering
\caption{Main retrieved and derived parameters of WASP-77A\,b from \bear compared with \citet{Smith2024}.}
\label{tab:w77}
\small
\begin{tabular}{lcccccc}
\toprule
& \multicolumn{2}{c}{IGRINS (HR)} & \multicolumn{2}{c}{NIRSpec (LR)} & \multicolumn{2}{c}{Joint (HR$+$LR)} \\
\cmidrule(lr){2-3}\cmidrule(lr){4-5}\cmidrule(lr){6-7}
Parameter & \citet{Smith2024} & \bear & \citet{Smith2024} & \bear & \citet{Smith2024} & \bear \\
\midrule
$\log_{10} X_\mathrm{H_2O}$ & $-3.97^{+0.11}_{-0.10}$ & $-3.74^{+0.14}_{-0.11}$ & $-3.80^{+0.34}_{-0.28}$ & $-3.90^{+0.28}_{-0.20}$ & $-4.02^{+0.07}_{-0.06}$ & $-3.69^{+0.15}_{-0.12}$ \\[3pt]
$\log_{10} X_\mathrm{CO}$ & $-3.81^{+0.19}_{-0.17}$ & $-3.65^{+0.20}_{-0.18}$ & $-3.73^{+0.55}_{-0.52}$ & $-4.16^{+0.48}_{-0.37}$ & $-3.91^{+0.13}_{-0.13}$ & $-3.76^{+0.22}_{-0.19}$ \\[3pt]
$\log_{10} X_\mathrm{CO_2}$ & --- & $<-4.41$ & $<-6.34$ & $<-6.68$ & $<-7.25$ & $-7.58^{+0.28}_{-0.57}$ \\[3pt]
$[\mathrm{^{13}CO/^{12}CO}]$ & $0.63^{+0.30}_{-0.34}$ & $0.83^{+0.27}_{-0.30}$ & $<1.78$ & $<2.41$ & $0.33^{+0.35}_{-0.71}$ & $0.60^{+0.33}_{-0.49}$ \\[3pt]
$T_0$ (K) & $1470^{+210}_{-370}$ & $1488^{+234}_{-480}$ & $1300^{+80}_{-50}$ & $1237^{+50}_{-49}$ & $1400^{+30}_{-40}$ & $1275^{+25}_{-26}$ \\[3pt]
$\delta\Kp$ (km\,s$^{-1}$) & $-1.26^{+0.85}_{-0.85}$ & $-0.45^{+0.90}_{-0.92}$ & --- & --- & $-1.29^{+0.75}_{-0.78}$ & $-0.34^{+0.89}_{-0.91}$ \\[3pt]
$\delta\Vsys$ (km\,s$^{-1}$) & $-5.27^{+0.51}_{-0.53}$ & $-6.60^{+0.60}_{-0.57}$ & --- & --- & $-5.27^{+0.48}_{-0.47}$ & $-6.66^{+0.60}_{-0.57}$ \\[3pt]
$\log_{10}\alpha$ (IGRINS) & $0.15^{+0.52}_{-0.53}$ & $0.25^{+0.74}_{-0.38}$ & --- & --- & $0.03^{+0.04}_{-0.04}$ & $-0.05^{+0.02}_{-0.02}$ \\[3pt]
$\log_{10}a_4$ (NIRSpec) & --- & --- & $0.01^{+0.02}_{-0.03}$ & fixed, 0 & $0.03^{+0.01}_{-0.02}$ & fixed, 0 \\
\midrule
C/O & $0.59^{+0.07}_{-0.07}$ & $0.57\pm0.07$ & $0.54^{+0.15}_{-0.16}$ & $0.36^{+0.13}_{-0.10}$ & $0.57^{+0.06}_{-0.06}$ & $0.47\pm0.06$ \\[3pt]
$[\mathrm{(C+O)/H}]$ & $-0.53^{+0.16}_{-0.13}$ & $-0.29^{+0.17}_{-0.15}$ & $-0.42^{+0.49}_{-0.42}$ & $-0.69^{+0.39}_{-0.27}$ & $-0.61^{+0.11}_{-0.10}$ & $-0.36^{+0.19}_{-0.16}$ \\
\bottomrule
\end{tabular}
\tablefoot{Medians with $1\sigma$ intervals (16th to 84th percentiles); upper limits are $3\sigma$ (99.73rd percentile), following \citet{Smith2024}. $\alpha$ is the IGRINS line contrast, called $a_2$ by \citet{Smith2024}, and $a_4$ their scale factor for the NIRSpec eclipse depths, which \bear does not fit. C/O and $[\mathrm{(C+O)/H}]$ are derived from the abundances of \ce{H2O}, CO, $^{13}$CO, \ce{CO2}, and \ce{CH4}, the latter relative to the solar values of \citet{Asplund2021}. \citet{Smith2024} counted $^{12}$CO only, included HCN, and used the solar values of \citet{Asplund2009}. With their definition, our C/O ratios are 0.555 (IGRINS) and 0.462 (joint) and our metallicities $-0.305$ and $-0.363$. Dashes mark parameters that a retrieval does not include. Appendix~\ref{app:comparison} lists all parameters.}
\end{table*}

\subsection{IGRINS-only retrieval}
\label{sec:w77}

\begin{figure}[ht]
  \centering
  \includegraphics[width=0.9\columnwidth]{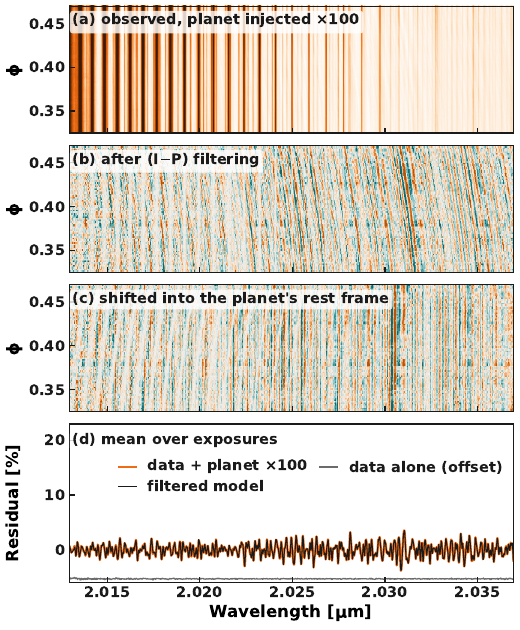}
  \vspace{-0.3cm}
  \caption{Recovery of a planet signal from IGRINS order 15 over the 79 exposures, with the best-fit model injected at 100 times its retrieved strength, shown over 2.013--2.037\,$\mu$m, the range covered by all exposures after the shift in (c). \emph{(a)} Observed spectra, each divided by its continuum, in the observatory frame, from deep absorption (dark) to the continuum (white); the stationary telluric and stellar lines form vertical stripes, and the planet lines form faint trails slanting to the left. \emph{(b)} The same after the $\IP$ filter, with the residual relative to the continuum in green (negative) and orange (positive), saturating at $\pm3.4\%$. \emph{(c)} Panel (b) after Doppler shifting, exposure by exposure, into the planet's rest frame with the retrieved $\Kp$ and $\Vsys$, which makes the planet trails vertical. \emph{(d)} Average of (c) over all exposures (orange), the planet model filtered and averaged in the same way (black), and the same average of the data without injection (gray, offset downward), in percent of the continuum.}
  \label{fig:w77_steps}
\end{figure}

We used the pre-eclipse IGRINS sequence of WASP-77A\,b taken on 2020 December 14 and first presented by \citet{Line2021}, a continuous 4.7\,h series of 79 spectra, each an A--B nod pair of 70\,s exposures, that covers orbital phases 0.325 to 0.470 (with $\phi=0.5$ at secondary eclipse) at $R\sim45\,000$ and a signal-to-noise ratio of 180 to 270 per resolution element. We started from the reduced and wavelength-calibrated spectra released with the example code of \citet{Line2021}, used the 44 orders of that data cube over 1.453--2.449\,$\mu$m, normalized each spectrum, and removed the first four singular components of each order, which capture the telluric and stellar variation with airmass; \citet{Line2021} found their results unchanged for four to ten removed components, and \citet{Smith2024} for four, six, or eight. The \bear observation file stores the normalized spectra, the singular vectors of each order, the orbital phases, the barycentric velocities, and the reference velocities $\Kp^\mathrm{ref}=192.06$\,km\,s$^{-1}$ and $\Vsys^\mathrm{ref}=1.6845$\,km\,s$^{-1}$ \citep{Line2021}, with the wavelengths left in the observatory frame. Figure~\ref{fig:w77_steps} follows one order through the pipeline of Sect.~\ref{sec:highres}, with the best-fit model injected at 100 times its retrieved strength. The planet lines are redshifted throughout the sequence ($v_\mathrm{rad}=+187$ to $+54$\,km\,s$^{-1}$), the $\IP$ filter removes the nearly stationary telluric and stellar lines, and in the planet's rest frame the planet lines add up over the exposures and match the model filtered in the same way. At its retrieved strength, the planet signal in this order is only 0.21 times the noise of the exposure average, so the detection requires all 44 orders.

For the retrieval, we computed the model on the high-resolution grid at $R=2.5\times10^5$ (186\,552 points) and broadened it with the \texttt{velocity\_broadening} module, holding $v\sin i=4.52$\,km\,s$^{-1}$ (synchronous rotation), $\epsilon=0$, $v_\mathrm{wind}=0$, and $\sigma_\mathrm{inst}=2.83$\,km\,s$^{-1}$ (corresponding to the IGRINS resolving power of about 45\,000) fixed as in \citet{Line2021}, whereas \citet{Smith2024} adopted $v\sin i=4.2$\,km\,s$^{-1}$. The stellar spectrum in the contrast was smoothed with a 200-pixel Gaussian (Sect.~\ref{sec:doppler}), and the phase function was not applied. As in \citet{Smith2024}, we enabled model filtering and re-injection and sampled $\alpha$ log-uniformly between 0.01 and 100, which selects the likelihood of Eq.~(\ref{eq:free_alpha}). The free $\alpha$ absorbs a mismatch in the overall line amplitude that is common to all exposures and that the one-dimensional model does not capture, for example from the approximate flux scale of the re-injected model or from averaging a non-uniform dayside over the observed phases \citep{Wardenier2025_3d}. With the velocities sampled as offsets from the reference values and the phase offset fixed at zero, the retrieval has 17 free parameters, which we sampled with 800 live points. Cross-correlating the filtered data with a template containing \ce{H2O} and CO yields a single peak near $\Kp=192$\,km\,s$^{-1}$ and $\delta\Vsys=-5$\,km\,s$^{-1}$ (Fig.~\ref{fig:w77_kpvsys}).
 
The retrieval constrains \ce{H2O} and CO and sets upper limits on all other gases (Table~\ref{tab:w77} and Fig.~\ref{fig:w77_summary}). We retrieve $\log_{10}X_\mathrm{H_2O}=-3.74^{+0.14}_{-0.11}$ and $\log_{10}X_\mathrm{CO}=-3.65^{+0.20}_{-0.18}$, within $1.5\sigma$ of \citet{Smith2024}, and a C/O ratio of $0.57\pm0.07$, which agrees with theirs to within $0.2\sigma$. We also reproduce their $^{13}$CO enrichment, which the Bayesian evidence prefers by a Bayes factor of about 50 to 90 (Appendix~\ref{app:13co}), and the negative systemic velocity offset \citep{Line2021,Smith2024}. Our $\delta\Vsys$ is lower than theirs by 1.3\,km\,s$^{-1}$, a difference of $1.7\sigma$. 

On its own, the line contrast $\log_{10}\alpha=0.25^{+0.74}_{-0.38}$ is poorly constrained, in agreement with \citet{Smith2024}, because it is almost fully degenerate with the temperature gradient. A steeper gradient deepens all lines by nearly the same factor as a larger $\alpha$. The posterior correlation coefficient of $\log_{10}\alpha$ is 0.98 with $\alpha_1$, which sets the temperature gradient of the upper layer where the line cores form, and 0.88 with $T_0$. The same degeneracy produces the curved IGRINS posterior in the $T_0$--$\alpha_1$ plane of Fig.~\ref{fig:w77_threeway_all}. The abundances, in contrast, are set by the relative depths of weak and strong lines and by the relative strengths of the \ce{H2O} and CO lines, and they are almost uncorrelated with $\alpha$ ($|r|\le0.03$). $T_0$ is defined at the top of the model atmosphere ($10^{-6}$\,bar), far above the pressures where the line cores form, and its IGRINS posterior mainly reflects the $\alpha$--gradient degeneracy and is consistent with that of \citet{Smith2024} (Table~\ref{tab:w77}). The data do constrain the temperature at 0.1\,bar, to $1816^{+86}_{-71}$\,K. Both species contribute near the CO band head, whereas \ce{H2O} alone accounts for the signal at shorter wavelengths (Fig.~\ref{fig:w77_spectrum}).

Further tests of the IGRINS retrieval are given in Appendices~\ref{app:robust} and \ref{app:injection}. The abundances change by at most 0.03\,dex when we leave the phase offset free, restrict the temperature profile to $P_3\ge P_1$, or replace \multinest by \nautilus. Without re-injection, the retrieval with the marginalized-$\alpha$ likelihood does not converge (Appendix~\ref{app:noreinject}). Synthetic planets with the composition of our joint fit, injected into the real data along four other velocity tracks, are recovered without a common offset in the abundances, but with a scatter that indicates underestimated posterior widths (Sect.~\ref{sec:disc_constraints}).

\subsection{NIRSpec-only retrieval}
\label{sec:w77_lr}

We applied the low-resolution secondary-eclipse model of \bear with its Gaussian likelihood to the NIRSpec/G395H eclipse spectrum of \citet{August2023}, which comprises 150 eclipse depths, 70 on NRS1 over 2.811--3.709\,$\mu$m and 80 on NRS2 over 3.835--5.164\,$\mu$m, in bins of about 13 and 17\,nm. \citet{Smith2024} fitted 160 points, ten more at 2.68--2.80\,$\mu$m. For each point, we adopted the upper of the asymmetric error bars of \citet{August2023} as its uncertainty. With the mean of the two error bars instead, every bounded parameter moves by at most $0.13\sigma$, and the $3\sigma$ limits become $\log_{10}X_\mathrm{CO_2}<-6.55$ and $[\mathrm{^{13}CO/^{12}CO}]<1.96$. The model was computed at a nominal resolving power of $10^5$ and integrated over the wavelength bins of the data. The retrieval includes \ce{H2O}, CO, \ce{CO2}, and $^{13}$CO and has ten free parameters, which we sampled with 2500 live points, as in \citet{Smith2024}. 
 
An eclipse-depth offset between the two detectors, which \citet{Smith2024} did not fit, is consistent with zero when we fitted it within $\pm200$\,ppm ($-50^{+36}_{-35}$\,ppm). However, the offset trades off against the planet-to-star flux ratio, correlates with the \ce{H2O} and CO abundances and with $T_0$ ($|r|=0.55$ to 0.72), and broadens the \ce{H2O} and CO posteriors by factors of 2.2 and 1.8, while the evidence slightly prefers a fixed offset ($\Delta\ln Z=0.37$). We therefore fixed the offset at zero in all retrievals that include NIRSpec. \citet{Smith2024} instead fitted a multiplicative scale factor $a_4$ for the NIRSpec eclipse depths, which \bear does not provide. In their NIRSpec retrieval it is consistent with unity ($\log_{10}a_4=0.01^{+0.02}_{-0.03}$), so it does not affect this comparison.
 
The retrieved composition and thermal structure agree with \citet{Smith2024} to within $1\sigma$ in every quantity they report (Table~\ref{tab:w77}). We retrieve $\log_{10}X_\mathrm{H_2O}=-3.90^{+0.28}_{-0.20}$ and $\log_{10}X_\mathrm{CO}=-4.16^{+0.48}_{-0.37}$, a C/O ratio of $0.36^{+0.13}_{-0.10}$, and a metallicity of $[\mathrm{(C+O)/H}]=-0.69^{+0.39}_{-0.27}$. As \citet{Smith2024} show, the low-resolution continuum constrains the thermal structure far better than the high-resolution data do, and we obtain $T_0=1237^{+50}_{-49}$\,K from NIRSpec, compared with $1488^{+234}_{-480}$\,K from IGRINS alone (Sect.~\ref{sec:w77}). Despite the different $T_0$, the temperature profiles of the two codes agree to within 30\,K between 0.01 and 1\,bar (Fig.~\ref{fig:tp_compare} and Table~\ref{tab:smith_full}). \ce{CO2} is bounded only from above, at $\log_{10}X_\mathrm{CO_2}<-6.68$, which is a tighter limit than the $<-6.34$ of \citet{Smith2024}. The $^{13}$CO ratio is likewise bounded only from above, with $<2.41$ at $3\sigma$ against their $<1.78$. The maximum-likelihood model reproduces the data with $\chi^2_\nu=0.65$ for $\nu=140$ (Fig.~\ref{fig:w77_nirspec}) and differs from the joint best fit of \citet{Smith2024} by at most 103\,ppm, or 1.1 times the data uncertainty (Fig.~\ref{fig:nirspec_compare}).

\subsection{Joint IGRINS+NIRSpec retrieval}
\label{sec:w77_joint}

The joint retrieval added the two NIRSpec files to the observation list of the IGRINS retrieval (Sect.~\ref{sec:joint_like}) and kept its free parameters and priors, with the NIRSpec detector offset fixed at zero. At each likelihood call, \bear computed the IGRINS model on the high-resolution grid ($R=2.5\times10^5$) and the NIRSpec model on the low-resolution grid ($R=10^5$) from the same atmosphere (Sect.~\ref{sec:hr_forward}).  \ce{CO2}, which has no detectable lines in the IGRINS orders, was kept out of the high-resolution opacities, and $\alpha$ scales only the IGRINS model. We sampled the posterior with 1000 live points.

\begin{figure}[htb]
  \centering
  \includegraphics[width=0.9\columnwidth]{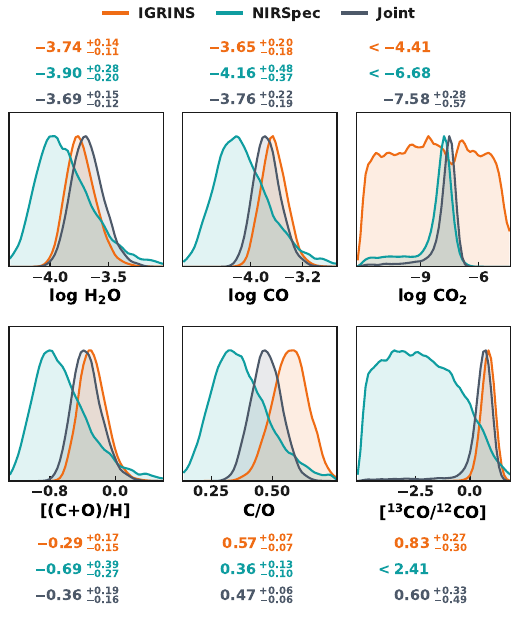}
  \vspace{-0.3cm}
  \caption{Retrieved composition of WASP-77A\,b from IGRINS (orange), NIRSpec (green), and both together (gray): marginalized posteriors of the \ce{H2O}, CO, and \ce{CO2} abundances, and of the derived metallicity $[\mathrm{(C+O)/H}]$, the C/O ratio, and the isotopologue ratio $[\mathrm{^{13}CO/^{12}CO}]$, each normalized to unit peak, with medians and $1\sigma$ intervals, or $3\sigma$ limits, printed in the color of each retrieval. The temperature profiles of the three retrievals are shown in Fig.~\ref{fig:tp_compare}.}
  \label{fig:w77_summary}
\end{figure}

\begin{figure*}[htb]
  \centering
  \includegraphics[width=0.8\textwidth]{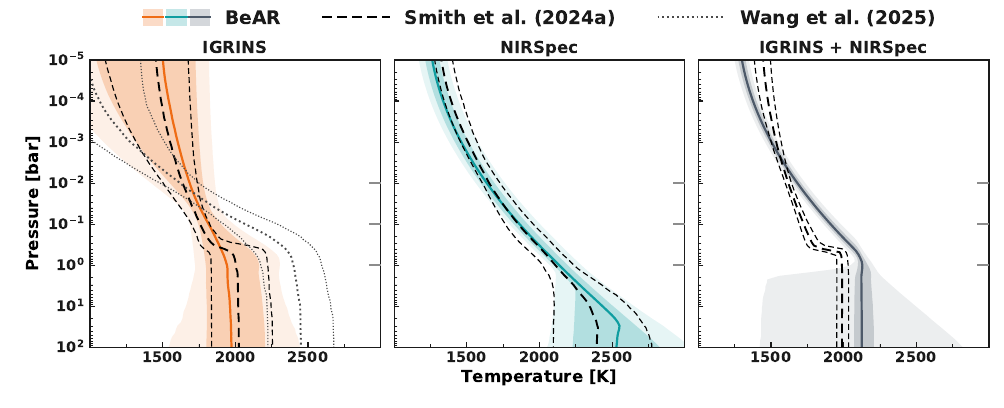}
  \vspace{-0.4cm}
  \caption{Retrieved temperature profiles of WASP-77A\,b from \bear (median as a solid line, $1\sigma$ and $2\sigma$ intervals shaded), compared with those of \citet{Smith2024} for the IGRINS, NIRSpec, and joint retrievals (black dashed) and with that of \citet{Wang2025} for IGRINS (gray dotted). For both, the thick line is the median and the thin lines bound the $1\sigma$ interval. These profiles are extracted from the vector graphics of \citet[Fig.~8]{Smith2024} and \citet[Fig.~13]{Wang2025}. The profile of \citet{Wang2025} is that of their PCA retrieval, with $\log_{10}X_\mathrm{H_2O}=-4.25$. Gray ticks mark the pressures of Table~\ref{tab:smith_full}.}
  \label{fig:tp_compare}
\end{figure*}

The joint posterior takes each constraint from the data set that provides it (Figs.~\ref{fig:w77_threeway}, \ref{fig:w77_summary}, and \ref{fig:tp_compare}). NIRSpec sets the thermal structure ($T_0=1275^{+25}_{-26}$\,K, $\alpha_1=0.456\pm0.011$) and the \ce{CO2} abundance ($\log_{10}X_\mathrm{CO_2}=-7.58^{+0.28}_{-0.57}$), IGRINS sets the isotopologue ratio ($[\mathrm{^{13}CO/^{12}CO}]=0.60^{+0.33}_{-0.49}$), and the \ce{H2O} and CO abundances ($-3.69^{+0.15}_{-0.12}$ and $-3.76^{+0.22}_{-0.19}$) are as precise as from IGRINS alone and about twice as precise as from NIRSpec alone. The C/O ratio of $0.47\pm0.06$ lies between the two individual values, and the metallicity of $[\mathrm{(C+O)/H}]=-0.36^{+0.19}_{-0.16}$ is $1.9\sigma$ below solar. NIRSpec constrains the temperature gradient, which breaks the $\alpha$--gradient degeneracy of the IGRINS retrieval and determines the line contrast, $\alpha=0.90\pm0.05$ ($\log_{10}\alpha=-0.05\pm0.02$). The abundances do not tighten further, since they are uncorrelated with $\alpha$ and the gradient in the IGRINS posterior. The joint solution fits both data sets nearly as well as the individual retrievals. Its maximum-likelihood sample has $\chi^2=93.8$ for the NIRSpec data, against 91.3 for the NIRSpec-only maximum-likelihood fit, and an IGRINS log-likelihood only 3.3 below the IGRINS-only maximum.

Our joint retrieval agrees with \citet{Smith2024} on a sub-solar metallicity and a near-solar C/O ratio (Table~\ref{tab:w77}). The narrower joint posteriors, however, expose differences that remain within the uncertainties of the individual retrievals. Among the abundances, CO and the isotopologue ratio agree with theirs to within $1\sigma$, and C/O and the metallicity to within $1.1\sigma$ and $1.3\sigma$. Only \ce{H2O} differs more, at $2.3\sigma$ above their value. The largest differences are in the thermal structure (Fig.~\ref{fig:tp_compare}), where our $T_0$ lies $2.6\sigma$ below theirs, with a correspondingly steeper gradient ($\alpha_1=0.456$ against 0.60). Our IGRINS line contrast lies $1.6\sigma$ below their $\log_{10}a_2$, and our $\delta\Vsys$ is lower than theirs by 1.4\,km\,s$^{-1}$ ($1.8\sigma$), whereas $\delta\Kp$ agrees to within $1\sigma$.
 
Two possible effects can contribute to these differences. First, the NIRSpec scale factor of \citet{Smith2024} departs from unity in their joint fit ($\log_{10}a_4=0.03^{+0.01}_{-0.02}$, about 7\%), so their NIRSpec flux level partly adjusts to the IGRINS lines, whereas in our setup the NIRSpec flux level fixes the thermal structure. Scaling our NIRSpec model by their factor makes the temperature gradient shallower ($\alpha_1=0.482\pm0.010$) and the atmosphere at 0.1\,bar 75\,K cooler, which raises our IGRINS line contrast to $\log_{10}\alpha=0.02\pm0.02$, in agreement with theirs, and C/O to $0.51\pm0.06$. The \ce{H2O} abundance ($-3.74^{+0.13}_{-0.12}$), the metallicity ($-0.37^{+0.18}_{-0.15}$), and $T_0$ ($1268^{+23}_{-24}$\,K) barely change. The scale factor therefore accounts for the difference in line contrast but not for those in \ce{H2O} and $T_0$. Second, the IGRINS-only \ce{H2O} abundances already differ by 0.23\,dex, and the joint fit inherits this difference at higher precision. The difference in $T_0$ says little about the photosphere because $T_0$ is defined at $10^{-6}$\,bar (Sect.~\ref{sec:w77_lr}), but the joint profiles also differ by 185\,K at 0.1\,bar. Although the joint profile shown by \citet{Smith2024} may come from a different run than their tabulated values (Fig.~\ref{fig:tp_compare}). Because our injection tests indicate that the high-resolution uncertainties are underestimated by a factor of 1.7 to 2 (Appendix~\ref{app:inj_calib}), these significances are likely overstated by a similar factor.

\section{Discussion}
\label{sec:discussion}

\subsection{What the high-resolution data constrain}
\label{sec:disc_constraints}

After filtering, dayside high-resolution data retain the depths of the planetary lines relative to the local pseudo-continuum but not the pseudo-continuum itself. Re-injection and a free $\alpha$ put model and data on a common scale (Sect.~\ref{sec:filtering}), yet in the WASP-77A\,b retrieval the absolute line contrast constrains only a combination of $\alpha$ and the temperature gradient (Sect.~\ref{sec:w77}). Independently of $\alpha$, the IGRINS data measure the relative depths of weak and strong lines and of the \ce{H2O} and CO lines \citep{Brogi2019}. IGRINS alone therefore constrains C/O ($\pm0.06$\,dex) more tightly than the metallicity ($\pm0.16$\,dex). It also constrains the temperature at 0.1\,bar to $\pm80$\,K, but not $T_0$ or the shape parameters of the profile.
 
A low-resolution spectrum breaks this degeneracy by fixing the temperature gradient, and in the joint retrieval $\alpha$ then becomes a test of the model rather than a nuisance parameter. Our retrieved value of $0.90\pm0.05$, or $1.06\pm0.06$ with the NIRSpec fluxes scaled as in \citet{Smith2024}, shows that the re-injected model reproduces the observed IGRINS line contrast to about 10\% once the thermal structure matches the NIRSpec continuum. We suggest that joint retrievals report this value as a consistency check between the high- and low-resolution forward models. Fixing the gradient did not sharpen our abundances, which are uncorrelated with it, whereas \citet{Smith2024} find their joint \ce{H2O} abundance to be 1.6 times more precise than from IGRINS alone. The gain from adding low-resolution data therefore depends on the forward model as well as on the data.

How precisely the high-resolution data constrain these parameters is a separate question. The cross-correlation likelihoods in use, including ours, assume uncorrelated noise, but the filtered IGRINS residuals are correlated, so in our injection tests the posteriors are too narrow by a factor of about two for abundance ratios and about 1.7 for the other parameters (Appendix~\ref{app:inj_calib}). Injections into the real residuals along many velocity tracks calibrate this effect directly, and we recommend them as a routine part of high-resolution retrievals.

\subsection{Differences between retrieval codes}
\label{sec:disc_codes}

Our IGRINS-only \ce{H2O} abundance lies 0.23\,dex above that of \citet{Smith2024}, and several differences between the two analyses may contribute. The two analyses differ in the tabulation and sampling of the opacities, in the radiative-transfer solver, and in whether the line broadening is applied in velocity space, which is required when the model grid is not uniform in velocity (Sect.~\ref{sec:broadening}). In addition, \chimera recomputes the filtering basis for every model, whereas \bear keeps it fixed \citep{Gibson2022}. With \poseidon, \citet{Wang2025} obtain $\log_{10}X_\mathrm{H_2O}=-4.33^{+0.10}_{-0.09}$ from the same data with a fixed planet radius and $-4.12\pm0.11$ with a free radius, both with PCA filtering but with fewer species. The three codes thus agree to within 0.6\,dex in the \ce{H2O} abundance despite independent forward models and setups, but this spread is still several times the statistical uncertainty of any single retrieval. The IGRINS temperature profiles of the three codes agree to within 70\,K at 0.01\,bar, but the \poseidon profile is 255\,K hotter than ours at 0.1\,bar ($1.5\sigma$ for the combined uncertainties; Fig.~\ref{fig:tp_compare}). Separating these effects requires comparing the forward models at fixed atmospheric parameters, as \citet{Barstow2020} did at low resolution. Open-source codes such as \bear, \poseidon, and \texttt{petitRADTRANS} now make such comparisons possible, and they are needed before high-resolution metallicities are interpreted at the level of their statistical uncertainties.

Comparisons of individual temperature-profile parameters are further complicated by the parameterization. The profile of \citet{Madhusudhan2009} is continuous only for $P_1\le P_3$, which the priors do not enforce, and 56 to 77\% of our posterior samples lie in the discontinuous region. Restricting the prior to $P_3\ge P_1$ leaves the abundances unchanged but changes the shape parameters substantially (Appendix~\ref{app:p3p1}), and the posteriors of \citet{Smith2024} indicate that \chimera also allows $P_3<P_1$. We therefore compare the profiles themselves and their temperatures at fixed pressures (Fig.~\ref{fig:tp_compare} and Table~\ref{tab:smith_full}), and recommend this practice, or parameterizations whose parameters are temperatures at fixed pressures, such as the PCHIP profile of \bear (Sect.~\ref{sec:bear_temperature}).

\section{Summary}
\label{sec:summary}

We have presented \bear, the open-source\footnote{\url{https://github.com/NewStrangeWorlds/BeAR}}, GPU-accelerated successor of \heliosr, and its new framework for high-resolution cross-correlation retrievals. The framework implements the likelihoods of \citet{Brogi2019} and \citet{Gibson2020,Gibson2022}, the fast model filtering of \citet{Gibson2022}, a re-injection of the model \citep{Line2021} that is compatible with this filter, and two velocity-broadening modules. High- and low-resolution likelihoods share one interface, so joint retrievals run within a single code, entirely on the GPU.

We repeated the IGRINS, NIRSpec, and joint retrievals of WASP-77A\,b by \citet{Smith2024}. We reproduce the abundances and C/O ratios of their individual retrievals to within $1.5\sigma$ and confirm a sub-solar metallicity, $[\mathrm{(C+O)/H}]=-0.36^{+0.19}_{-0.16}$, with a near-solar C/O ratio of $0.47\pm0.06$. In the joint fit, NIRSpec constrains the thermal structure and with it the IGRINS line contrast, whereas IGRINS sets the isotopologue ratio. The remaining differences in \ce{H2O} ($2.3\sigma$) and $T_0$ ($2.6\sigma$) may reflect differences between the forward models. Re-injection is essential for this emission retrieval, and injection--recovery tests show that the high-resolution posteriors are too narrow by up to a factor of two. A full IGRINS retrieval takes 27 to 42\,min on a single NVIDIA GeForce RTX 5080 and about 14\,min on eight RTX 4090 GPUs.

These results suggest a few practices for high-resolution and joint retrievals. We recommend stating the line-broadening treatment and the spectral grid, comparing temperature profiles at fixed pressures rather than through their parameters, reporting the line contrast of joint fits as a consistency check, comparing evidences only between runs with the same sampler, and calibrating the posterior widths with injections into the real residuals.

\begin{acknowledgements}
This work was supported by the Swiss National Science Foundation under grant 200021-231596. H.J.H. additionally acknowledges support of grants from eSSENCE (grant number eSSENCE@LU 9:3), the Swedish National Research Council (project number 2023-05307), The Royal Physiographic Society, and The Crafoord foundation. We thank M.~R.~Line for making the reduced IGRINS spectra of WASP-77A\,b and the \chimera example code publicly available \citep{Line2021}, and P.~C.~B.~Smith for releasing the model spectra \citep{Smith2023zenodo}. This publication makes use of the Data \& Analysis Center for Exoplanets (DACE), a platform of the Swiss National Centre of Competence in Research PlanetS, which hosts the \heliosk opacity database. This research made use of \fastchem, \multinest, and \nautilus. Part of this work was supported by Haku, Fluffy and Bärli and we shall be lastingly grateful.
\end{acknowledgements}

\bibliographystyle{aa} 
\bibliography{references}

\begin{appendix}
\nolinenumbers
\section{Additional figures}
\label{app:figures}
\begin{figure}[htb]
  \centering
  \includegraphics[width=0.8\columnwidth]{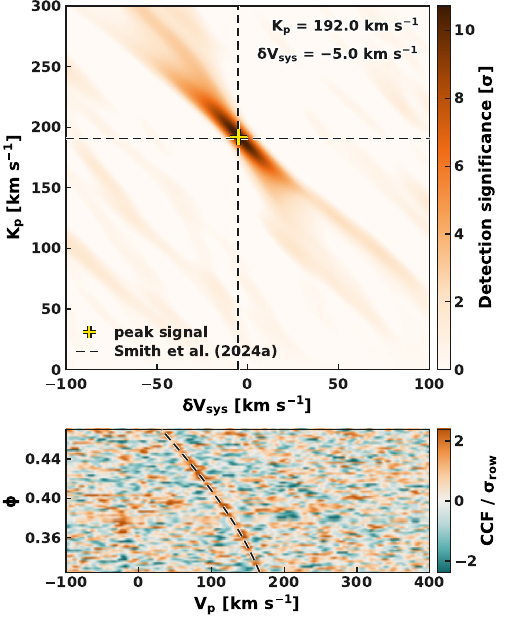}
  \caption{Cross-correlation detection of WASP-77A\,b in emission with IGRINS, using a template containing \ce{H2O} and CO. \emph{Upper:} detection significance as a function of $\Kp$ and $\delta\Vsys$, obtained by phase-folding the cross-correlation matrix below along each trial $\Kp$ track and normalizing by the scatter of the map away from the peak. Dashed lines indicate the values of \citet{Smith2024}, and the cross marks the peak. \emph{Lower:} cross-correlation divided by its standard deviation in each exposure, as a function of orbital phase $\phi$ and planet radial velocity $V_\mathrm{p}$ in the stellar rest frame. The planet's trail is the diagonal feature along the dashed track.}
  \label{fig:w77_kpvsys}
\end{figure}
\begin{figure}
  \includegraphics[width=\columnwidth]{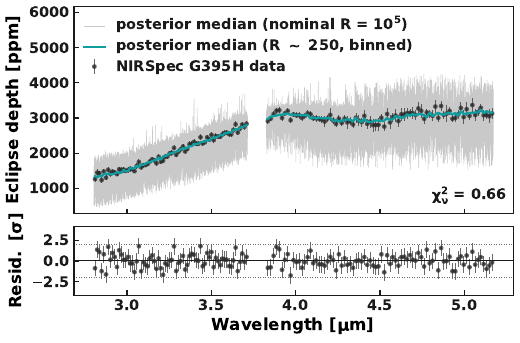}
  \caption{NIRSpec/G395H secondary-eclipse spectrum of WASP-77A\,b with the posterior-median \bear model. Points with error bars are the observed eclipse depths. The gray curve is the posterior median of the models on the native model grid (nominal $R=10^5$), and the green curve is the posterior median of the same models binned to the resolution of the data ($R\sim250$). The binned median gives $\chi^2_\nu=0.66$. The gap at 3.7--3.8\,$\mu$m separates NRS1 and NRS2. The lower panel shows the residuals in units of the observational uncertainty, with dotted lines marking the $\pm2\sigma$ levels.}
  \label{fig:w77_nirspec}
\end{figure}

\begin{figure}[htb]
  \centering
  \includegraphics[width=\columnwidth]{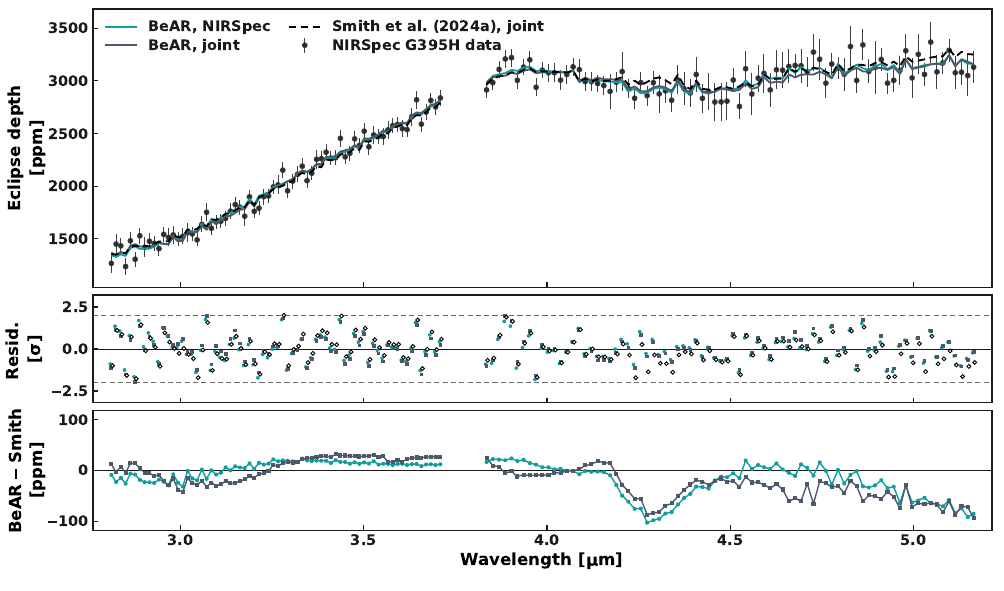}
  \caption{NIRSpec/G395H eclipse spectrum of WASP-77A\,b with the maximum-likelihood models of our NIRSpec-only (green) and joint (gray) retrievals and the joint best-fit model of \citet{Smith2024} (black dashed). Their model is taken from their data release \citep{Smith2023zenodo} and shown in their Fig.~9. Its file name and a fit of its overall scale indicate that it includes their NIRSpec scale factor. \emph{Upper:} All models are binned to the data as in the \bear likelihood and give $\chi^2=91.3$, 93.8, and 105.3 for the 150 data points. \emph{Middle:} residuals of our NIRSpec-only and joint models (filled circles and squares) and of the model of \citet{Smith2024} (open diamonds), in units of the observational uncertainty. \emph{Lower:} difference between our models and that of \citet{Smith2024}.}
  \label{fig:nirspec_compare}
\end{figure}

 \begin{figure}
  \centering
  \includegraphics[width=\columnwidth]
  {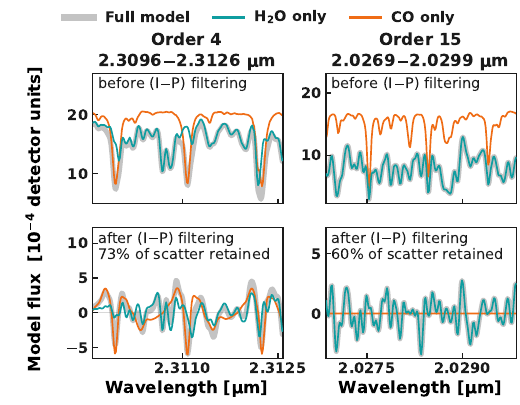}
  \caption{Best-fit IGRINS model and its decomposition into species, before (upper) and after (lower) the $\IP$ filter, in 3\,nm windows of order 4 (2.3096--2.3126\,$\mu$m), where CO and \ce{H2O} contribute comparably, and of order 15 (2.0269--2.0299\,$\mu$m), where \ce{H2O} accounts for the model on its own. Gray is the full model. Green shows the model with every absorber but \ce{H2O} suppressed, and orange the model with every absorber but CO (with $^{13}$CO) suppressed. The upper panels show the model as the likelihood constructs it, including the re-injection scaling, so the vertical scale is in detector units. The exposure shown is at $\phi=0.399$, mid-way through the observation sequence. The annotation gives the fraction of the model's scatter in the window that survives the filter; over the full orders it is 38\% and 29\%.}
  \label{fig:w77_spectrum}
\end{figure}

\begin{figure}[htb]
  \includegraphics[width=\columnwidth]{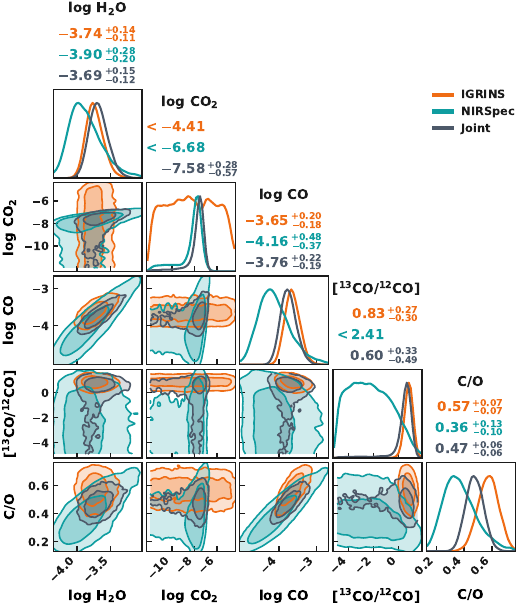}
  \caption{Posterior distributions of the IGRINS (orange), NIRSpec (green), and joint (gray) retrievals for the abundances, the isotopologue ratio, and the derived C/O ratio. Contours enclose 68\% and 95\% of the samples; the diagonal panels give each retrieval's median and $1\sigma$ interval, or its $3\sigma$ limit where the posterior reaches a prior edge. All shared parameters are shown in Fig.~\ref{fig:w77_threeway_all}.}
  \label{fig:w77_threeway}
\end{figure}
 
\begin{figure*}[htb]
  \centering
  \includegraphics[width=\textwidth]{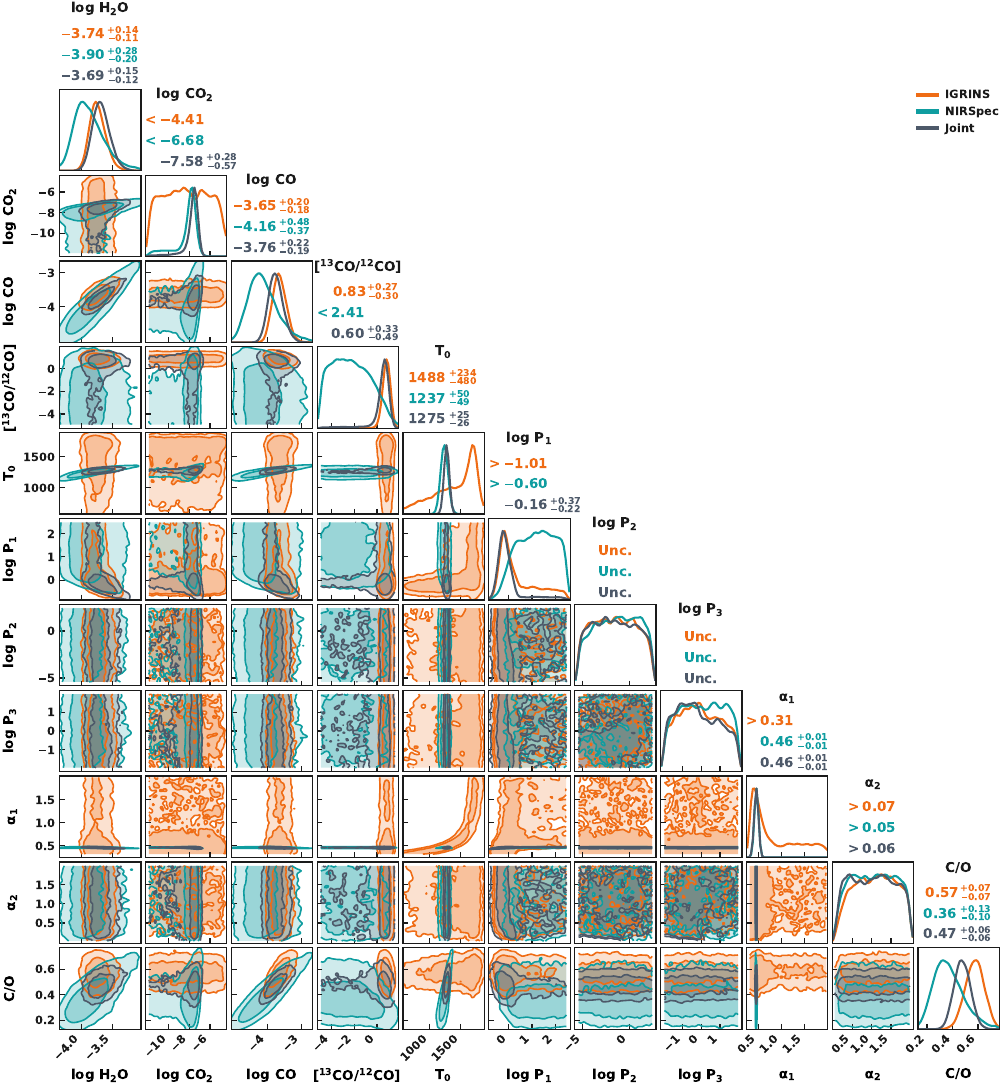}
  \caption{As Fig.~\ref{fig:w77_threeway}, for all parameters shared by the IGRINS (orange), NIRSpec (green), and joint (gray) retrievals, including the six parameters of the temperature profile. The curved IGRINS posterior in the $T_0$--$\alpha_1$ plane reflects the degeneracy of both parameters with the line contrast (Sect.~\ref{sec:w77}), which NIRSpec breaks.}
  \label{fig:w77_threeway_all}
\end{figure*}
 
\clearpage

\section{Robustness tests of the IGRINS retrieval}
\label{app:robust}

Table~\ref{tab:robust} summarizes retrievals of the IGRINS data that each vary one aspect of the benchmark setup of Sect.~\ref{sec:w77} or of the reduced model.

\subsection{Choice of sampler}
\label{app:sampler}

Repeating the benchmark with \nautilus \citep{Lange2023}, with 800 live points and an otherwise identical configuration, reproduces all 17 posterior medians of the \multinest run to within $0.22\sigma$. The log-evidences, however, differ by up to 4.8 between the samplers (381.89 with nested sampling and 377.06 with importance nested sampling \citep{Feroz2019} in \multinest, and 377.81 with \nautilus), whereas evidence differences between models computed with the same sampler agree to within 0.6, so model comparisons should use a single sampler throughout. \nautilus required 176\,100 likelihood evaluations, which is 3.3 times fewer than \multinest, but finished only 1.5 times faster because the training of the neural networks with which \nautilus constructs its bounds takes up a larger share of its shorter wall time.

\subsection{Evidence for $^{13}$CO and the phase offset}
\label{app:13co}

We tested the evidence for $^{13}$CO and the phase offset with a reduced model that contains only \ce{H2O}, CO, and $^{13}$CO, with $^{13}$CO retrieved as its own mixing ratio. With $\delta\phi$ free, removing $^{13}$CO lowers the log-evidence by 4.53 with \multinest, by 3.95 with its importance-nested-sampling estimate, and by 4.20 with \nautilus. These values correspond to Bayes factors of about 50 to 90 in favor of $^{13}$CO. The calibration of \citet{Sellke2001}, as tabulated by \citet{Benneke2013}, would translate them into 3.3 to 3.5$\sigma$, but \citet{Thorngren2026} show that this conversion is not valid and generally overstates the significance. We therefore quote the Bayes factors, which at equal prior odds give the model without $^{13}$CO a probability of 0.01 to 0.02. A ratio $\mathrm{^{12}CO/^{13}CO}\approx13$, seven times below the terrestrial value, would be unusual for a hot Jupiter. The $^{13}$CO lines also lie close to those of $^{12}$CO and can absorb small mismatches in the modeled CO lines. We therefore regard the enrichment as tentative until it is confirmed independently.
 
Leaving $\delta\phi$ free within $\pm0.01$ returns $\delta\phi=-0.003^{+0.004}_{-0.003}$, consistent with zero, changes the abundances by at most 0.01\,dex, and changes the evidence by at most 0.5 ($-0.25$ with nested sampling and $-0.50$ with importance nested sampling). It does, however, widen the $\delta\Kp$ and $\delta\Vsys$ posteriors by factors of 3.9 and more than 9.2, the latter limited by the prior, so we fixed it in the benchmark.

\subsection{Ordering of the temperature-profile pressures}
\label{app:p3p1}

The temperature profile of \citet{Madhusudhan2009} is continuous only for $P_1\le P_3$. The priors of \citet{Smith2024}, which we adopted for the benchmark, sample $P_1$, $P_2$, and $P_3$ independently and do not enforce this ordering. For $P_3<P_1$, the \bear implementation follows the upper layer down to $P_1$ and then steps directly to the isothermal $T_3$, so the profile is discontinuous at $P_1$. This occurs in 62\% of the benchmark posterior samples. To test whether these discontinuous profiles affect our results, we restricted the posterior to samples with $P_3\ge P_1$, which is equivalent to adding this condition to the prior. This restriction shifts the abundances by at most 0.03\,dex, C/O by less than 0.01, and $T_0$ by $-130$\,K. The retrieved composition therefore does not depend on the ordering, whereas the shape parameters of the profile do (Sect.~\ref{sec:disc_codes}).

\subsection{Retrieval without re-injection}
\label{app:noreinject}

The marginalized-$\alpha$ likelihood is the simplest option for an emission retrieval, because it needs no prior on $\alpha$ and no pre-scaling of the model (Sect.~\ref{sec:like}). We therefore repeated the reduced-model retrieval of Appendix~\ref{app:13co}, with $\delta\phi$ free, without re-injection and with this likelihood, to test whether re-injection is required. The retrieval failed in this setup. The sampler collapsed onto models with every abundance at the prior floor, $T_0$ at its lower bound, and $\delta\Kp$ and $\delta\phi$ at the edges of their priors, and the run had not converged after 2\,h, with a nominal $\ln Z\approx1.06\times10^4$. 

To identify the cause, we evaluated the likelihood of this configuration directly at three points. The best-fit physical model of the benchmark reaches $\ln\mathcal{L}=3312$, the same model with all abundances lowered by eight orders of magnitude reaches 3629, and the collapsed point reaches 10\,682. This likelihood rescales every order and exposure independently and with either sign, so it rewards any model structure that correlates with residual systematics in the filtered data, including the residual broadband structure of nearly featureless model spectra, as anticipated in Sect.~\ref{sec:like}. Re-injection with a single free $\alpha$ removes this per-Echelle-order, per-exposure scaling, and we therefore used it in the benchmark and recommend it for emission retrievals with high-resolution data.

\begin{table*}[htb]
\centering
\caption{Robustness tests of the IGRINS retrieval.}
\label{tab:robust}
\small
\begin{tabular}{lcccccccc}
\toprule
Retrieval & $N_\mathrm{free}$ & $\log_{10} X_\mathrm{H_2O}$ & $\log_{10} X_\mathrm{CO}$ & C/O & $[\mathrm{(C+O)/H}]$ & $\log_{10}\alpha$ & $\delta\Vsys$ (km\,s$^{-1}$) & $\ln Z$ \\
\midrule
Benchmark (\multinest) & 17 & $-3.74^{+0.14}_{-0.11}$ & $-3.65^{+0.20}_{-0.18}$ & $0.57\pm0.07$ & $-0.29^{+0.17}_{-0.15}$ & $0.25^{+0.74}_{-0.38}$ & $-6.60^{+0.60}_{-0.57}$ & 381.89 \\[3pt]
Benchmark (\nautilus) & 17 & $-3.73^{+0.15}_{-0.12}$ & $-3.63^{+0.23}_{-0.19}$ & $0.58\pm0.07$ & $-0.28^{+0.19}_{-0.15}$ & $0.35^{+0.65}_{-0.43}$ & $-6.60^{+0.61}_{-0.58}$ & 377.81 \\[3pt]
Benchmark, $P_3\ge P_1$ & 17 & $-3.73^{+0.14}_{-0.11}$ & $-3.62^{+0.21}_{-0.18}$ & $0.58\pm0.07$ & $-0.27^{+0.18}_{-0.15}$ & $0.10^{+0.81}_{-0.27}$ & $-6.60^{+0.59}_{-0.58}$ & -- \\[3pt]
Reduced, $\delta\phi$ free & 13 & $-3.77^{+0.14}_{-0.12}$ & $-3.68^{+0.21}_{-0.20}$ & $0.57\pm0.08$ & $-0.33^{+0.18}_{-0.16}$ & $-0.14^{+0.32}_{-0.12}$ & $-11.1^{+6.0}_{-5.2}$ & 385.06 \\[3pt]
Reduced, $\delta\phi=0$ & 12 & $-3.76^{+0.14}_{-0.12}$ & $-3.68^{+0.21}_{-0.19}$ & $0.57^{+0.07}_{-0.08}$ & $-0.32^{+0.18}_{-0.16}$ & $-0.15^{+0.31}_{-0.12}$ & $-6.61^{+0.62}_{-0.58}$ & 385.30 \\[3pt]
Reduced, no $^{13}$CO & 12 & $-3.79^{+0.13}_{-0.11}$ & $-3.70^{+0.20}_{-0.19}$ & $0.55\pm0.08$ & $-0.37^{+0.17}_{-0.15}$ & $-0.16^{+0.33}_{-0.12}$ & $-12.1^{+5.6}_{-4.8}$ & 380.53 \\
\bottomrule
\end{tabular}
\tablefoot{The benchmark is the retrieval of Sect.~\ref{sec:w77}. ``$P_3\ge P_1$'' is its posterior restricted to samples with $P_3\ge P_1$, equivalent to a retrieval with this condition added to the prior. The reduced model contains \ce{H2O}, CO, and $^{13}$CO only, with $^{13}$CO retrieved as its own mixing ratio and $\delta\phi$ free within $\pm0.01$ unless stated otherwise. Its priors are log-U$[10^{-12},0.1]$ for the mixing ratios, log-U$[0.1,10]$ for $\alpha$, U$[0.02,1.04]$ for $\alpha_1$ and $\alpha_2$, and U$[-25,25]$\,km\,s$^{-1}$ for $\delta\Kp$, identical in all three runs. The $\delta\Vsys$ posteriors with free $\delta\phi$ reach the edge of their prior. $\ln Z$ is the \multinest nested-sampling evidence except in the \nautilus row. All rows use the same data and likelihood, so their evidences can be compared, but only between runs with the same sampler. The restricted posterior has no evidence of its own.}
\end{table*}

\section{Injection--recovery tests}
\label{app:injection}


To test whether the retrieval recovers a known atmosphere from these data, we injected synthetic planets into the real IGRINS spectra following \citet[][Sect.~4]{Wang2025}. We computed each planet spectrum with the benchmark forward model and broadening, Doppler-shifted it to a velocity track well separated from that of the real planet, and multiplied it into the observed fluxes of every order and exposure, $D_\mathrm{inj}=D\,(1+m_\mathrm{hr})$, with $m_\mathrm{hr}$ the planet-to-star contrast of Sect.~\ref{sec:hr_forward}. We then recomputed the four singular vectors of each order from the injected data, so that the injected signal passes through the detrending as a real one would. Each data set was retrieved with the benchmark setup and priors, with the reference velocities set to the injected ones.

\subsection{Results}
\label{app:inj_results}

\begin{figure*}[htb]
  \centering
  \includegraphics[width=0.45\textwidth]{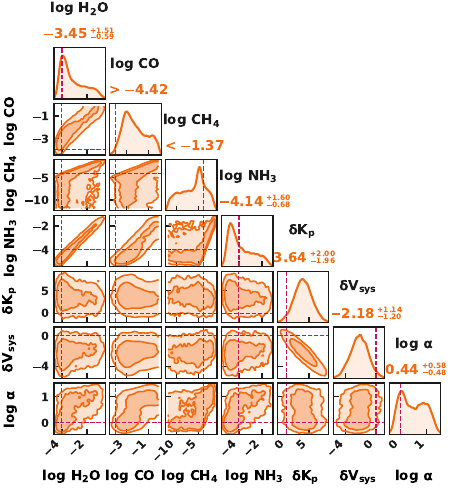}
  \hfill
  \includegraphics[width=0.45\textwidth]{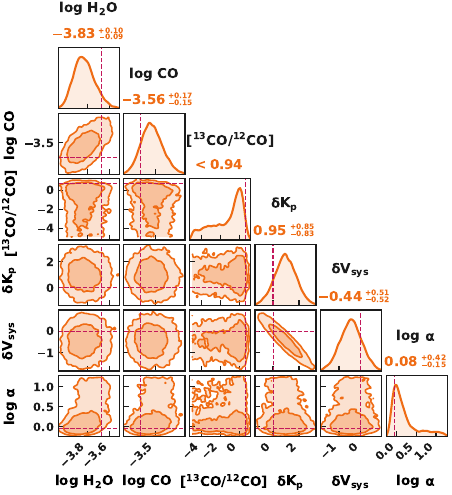}
  \caption{Posteriors of two injection--recovery retrievals on the IGRINS data, with the injected values marked by the red dashed lines (contours enclose 68\% and 95\% of the posterior). \emph{Left:} Composition of \citet{Wang2025}, with \ce{H2O}, CO, \ce{CH4}, and \ce{NH3} at $\log_{10}X=-4$, injected at $(\Kp,\Vsys)=(-200,-20)$\,km\,s$^{-1}$. \emph{Right:} Maximum-likelihood atmosphere of the joint fit ($\log_{10}X_\mathrm{H_2O}=-3.67$, $\log_{10}X_\mathrm{CO}=-3.73$, $[\mathrm{^{13}CO/^{12}CO}]=0.66$, $\alpha=0.90$), injected at the same velocities. $\delta\Kp$ and $\delta\Vsys$ are measured from the injected velocities. Values above the diagonal are medians with $1\sigma$ intervals, or $3\sigma$ limits where the posterior reaches a prior edge (Appendix~\ref{app:comparison}).}
  \label{fig:w77_injection_corner}
\end{figure*}

We first adopted the composition of the injection test of \citet{Wang2025}, with \ce{H2O}, CO, \ce{CH4}, and \ce{NH3} at $\log_{10}X=-4$. We injected it at $\Kp=-200$ and $\Vsys=-20$\,km\,s$^{-1}$ on the thermal profile of our joint fit, since the profile of \citet{Wang2025} is anchored at 0.01\,bar instead of at the top of the atmosphere. The strong \ce{CH4} and \ce{NH3} opacity across much of the $H$ and $K$ bands lowers the line contrast, so this signal is weak ($\ln Z=40$, against $\ln Z=382$ for the real planet). All injected values lie within $1.9\sigma$ of the retrieved medians, but the absolute abundances are degenerate along a common ridge, as \citet{Wang2025} also find (Fig.~\ref{fig:w77_injection_corner}, left).

\begin{figure*}[htb]
  \centering
  \includegraphics[width=0.96\textwidth]{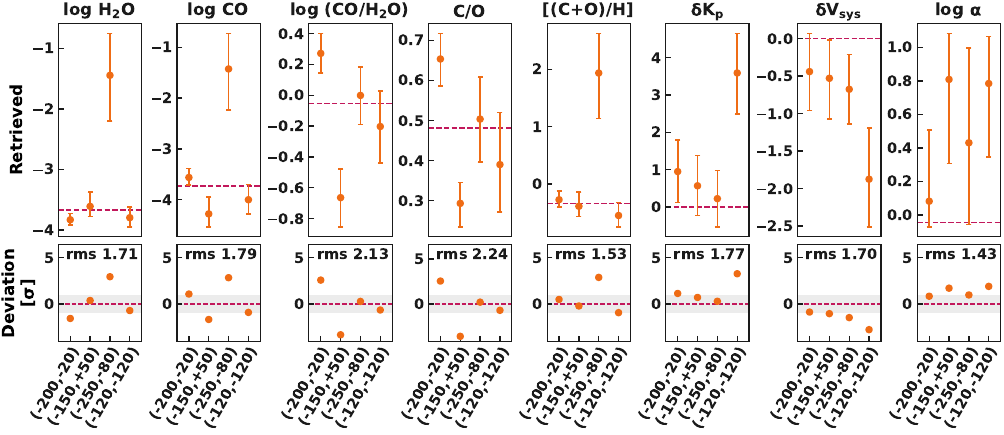}
  \caption{Injection--recovery tests at the signal strength of the real planet. \emph{Upper:} retrieved medians and $1\sigma$ intervals for four injections of the joint-fit atmosphere of WASP-77A\,b along different velocity tracks (labeled by the injected $\Kp$ and $\Vsys$ in km\,s$^{-1}$); the dashed lines mark the injected values. \emph{Lower:} deviations from the injected values in units of the posterior width, with their rms over the four injections; the gray band marks $\pm1\sigma$. The four injections share the same atmosphere and differ only in the velocity track, which overlaps a different part of the residual noise. For calibrated posteriors, the rms deviation would be one.}
  \label{fig:w77_injection}
\end{figure*}

To test the retrieval at the signal strength of the real planet, we then injected the maximum-likelihood atmosphere of our joint fit among the samples with $P_3\ge P_1$ ($\log_{10}X_\mathrm{H_2O}=-3.67$, $\log_{10}X_\mathrm{CO}=-3.73$, $[\mathrm{^{13}CO/^{12}CO}]=0.66$, its thermal profile, and all other absorbers at the prior floor), scaled by its line contrast of $\alpha=0.90$, along four velocity tracks, $(\Kp,\Vsys)=(-200,-20)$, $(-150,+50)$, $(-250,-80)$, and $(-120,-120)$\,km\,s$^{-1}$ (Table~\ref{tab:injection} and Fig.~\ref{fig:w77_injection}). Each track overlaps a different part of the residual noise. The right panel of Fig.~\ref{fig:w77_injection_corner} shows the full posterior of the first track. It recovers the injected values, but with a $\Kp$--$\Vsys$ degeneracy, an extended tail of $\alpha$ towards larger values, and only an upper limit on the isotopologue ratio. The retrieved values scatter about the injected ones without a common offset in the abundances. The deviations of $\log_{10}(\mathrm{CO/H_2O})$ are $+2.6$, $-3.3$, $+0.3$, and $-0.6\sigma$, and the metallicity is recovered to within $1\sigma$ in three of the four cases.

\begin{figure*}[htb]
  \centering
  \includegraphics[width=0.45\textwidth]{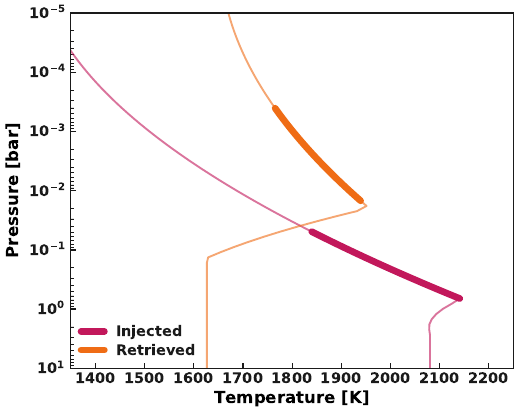}
  \caption{Thermal profiles of the injected atmosphere (red) and of the maximum-likelihood sample of the injection at $(\Kp,\Vsys)=(-250,-80)$\,km\,s$^{-1}$ (orange; $\log_{10}X_\mathrm{H_2O}=-1.51$, $\log_{10}X_\mathrm{CO}=-1.55$). The thick part of each profile marks its photosphere. For every pixel of the IGRINS data, we take the pressure at which the contribution function (the fraction of the emergent flux emitted by each layer) peaks, and the thick part spans the 5th to 95th percentiles of these pressures. Outside these ranges, the data do not constrain the profiles.}
  \label{fig:w77_injection_saturation}
\end{figure*}

In one case, $(-250,-80)$\,km\,s$^{-1}$, only the ratio of the abundances is recovered (to within $0.3\sigma$), while the absolute abundances rise to $\log_{10}X\approx-1.4$, with $3\sigma$ lower limits of $-3.05$ for \ce{H2O} and $-3.29$ for CO. At such abundances, \ce{H2O} and CO dominate the opacity at all IGRINS wavelengths, including between the lines, so the photosphere moves from about 0.3\,bar to about 0.005\,bar, and the retrieved thermal profile follows it (Fig.~\ref{fig:w77_injection_saturation}). Across this higher photosphere, the temperature changes by 170 instead of 300\,K, so that at equal $\alpha$ the filtered lines are only 0.37 times as deep as the injected ones. A larger line contrast compensates for this, with $\log_{10}\alpha=0.39$ for the maximum-likelihood sample against the injected $-0.05$. After filtering, the two models are nearly identical (correlation coefficient 0.98), because the data retain only the line depths relative to the local pseudo-continuum (Sect.~\ref{sec:disc_constraints}), and here that pseudo-continuum is itself formed by \ce{H2O} and CO.

\subsection{Posterior widths and systematic offsets}
\label{app:inj_calib}

Through our injection tests, we find that the scatter of the retrieved values exceeds their quoted uncertainties. In units of the posterior widths, the deviations of C/O and of $\log_{10}(\mathrm{CO/H_2O})$ have an rms of 2.2 and 2.1, and those of the individual abundances, the metallicity, and the velocities have an rms of 1.5 to 1.8. With four injections, these rms values are uncertain by about a third, but they indicate that the posterior widths underestimate the true uncertainties by a factor of about two for abundance ratios and of about 1.7 for the other parameters. This is consistent with correlated residual noise, which the high-resolution likelihood treats as uncorrelated, as do those of \citet{Brogi2019} and \citet{Gibson2020}. The scatter of the cross-correlation function away from the planet exceeds its white-noise expectation by a factor of 1.2 to 1.7 on these data. 

Three parameters also deviate in the same direction in all four injections. The $\log_{10}\alpha$ deviations of $+0.1$ to $+0.9$ arise from the $\alpha$--gradient degeneracy, along which the sampler favors a larger $\alpha$ with a shallower gradient. The deviations of $\delta\Kp$ ($+0.2$ to $+3.6$\,km\,s$^{-1}$) and $\delta\Vsys$ ($-0.4$ to $-1.9$\,km\,s$^{-1}$) reflect a single deviation. The observations cover a short orbital arc, so $\Kp$ and $\Vsys$ are degenerate (correlation coefficient $-0.95$) and shift together along this degeneracy. The well-constrained quantity is the planet's velocity at the middle of the sequence, $\delta v_\mathrm{mid}=\delta\Vsys+\sin(2\pi\phi_\mathrm{mid})\,\delta\Kp$, where $\phi_\mathrm{mid}=0.398$ is the mean orbital phase. It is constrained to $\pm0.14$ to $\pm0.19$\,km\,s$^{-1}$ and deviates by $+0.13$, $-0.20$, $-0.54$, and $+0.28$\,km\,s$^{-1}$ in the four injections, without a common offset.

To test whether these deviations come from the residual noise of the data or from the injection or the retrieval, we repeated the injections with the noise reduced. For each order, the four-component reconstruction $\mathbf{P}\mathbf{D}$ holds the stellar, telluric, and instrumental variation, and the residuals $\mathbf{D}-\mathbf{P}\mathbf{D}=\IP\mathbf{D}$ (Sect.~\ref{sec:filtering}) hold the noise, any remaining systematics, and the real planet signal. We replaced the data by $\mathbf{P}\mathbf{D}+0.1\,\IP\mathbf{D}$, which keeps the large-scale structure and reduces the residuals by a factor of ten, injected the same signals, and otherwise followed the procedure above. We then located the maximum of the likelihood at the injected atmosphere. An error in the injection or the likelihood, such as a velocity convention or an interpolation error, does not depend on the noise and would leave the offsets unchanged. Offsets caused by the noise should shrink in proportion, by a factor of 0.1. The offsets in $\delta\Kp$, $\delta\Vsys$, and $\delta v_\mathrm{mid}$ shrink by factors of 0.07 to 0.12, so they arise from the residual noise of the data.

\begin{table*}[htb]
\centering
\caption{Injection--recovery tests at the signal strength of the real planet.}
\label{tab:injection}
\footnotesize
\begin{tabular}{lcccccccc}
\toprule
$(\Kp,\Vsys)$ (km\,s$^{-1}$) & $\ln Z$ & $\log_{10} X_\mathrm{H_2O}$ & $\log_{10} X_\mathrm{CO}$ & $\log_{10}(\mathrm{CO/H_2O})$ & C/O & $[\mathrm{(C+O)/H}]$ & $\delta\Kp$ (km\,s$^{-1}$) & $\delta\Vsys$ (km\,s$^{-1}$) \\
\midrule
Injected & -- & $-3.67$ & $-3.73$ & $-0.05$ & $0.48$ & $-0.34$ & $0$ & $0$ \\
\midrule
$(-200,-20)$ & 421.0 & $-3.83^{+0.10}_{-0.09}$ & $-3.56^{+0.17}_{-0.15}$ & $0.27\pm0.13$ & $0.65^{+0.06}_{-0.07}$ & $-0.27^{+0.15}_{-0.13}$ & $0.95^{+0.85}_{-0.83}$ & $-0.44^{+0.51}_{-0.52}$ \\[3pt]
$(-150,+50)$ & 345.9 & $-3.61^{+0.24}_{-0.17}$ & $-4.28^{+0.33}_{-0.25}$ & $-0.66^{+0.18}_{-0.19}$ & $0.29^{+0.05}_{-0.06}$ & $-0.39^{+0.25}_{-0.18}$ & $0.57^{+0.81}_{-0.80}$ & $-0.53^{+0.51}_{-0.55}$ \\[3pt]
$(-250,-80)$ & 450.5 & $-1.45^{+0.69}_{-0.76}$ & $-1.43^{+0.70}_{-0.81}$ & $0.00\pm0.19$ & $0.50^{+0.10}_{-0.11}$ & $1.94^{+0.69}_{-0.79}$ & $0.22\pm0.75$ & $-0.68\pm0.46$ \\[3pt]
$(-120,-120)$ & 240.0 & $-3.80^{+0.18}_{-0.15}$ & $-4.00^{+0.30}_{-0.28}$ & $-0.20^{+0.23}_{-0.24}$ & $0.39^{+0.13}_{-0.12}$ & $-0.55^{+0.23}_{-0.20}$ & $3.59^{+1.06}_{-1.10}$ & $-1.88^{+0.68}_{-0.63}$ \\
\midrule
rms deviation ($\sigma$) & & 1.7 & 1.8 & 2.1 & 2.2 & 1.5 & 1.8 & 1.7 \\
\bottomrule
\end{tabular}
\tablefoot{The injected atmosphere is the maximum-likelihood sample of the joint fit with $P_3\ge P_1$ (\ce{H2O}, CO, $^{13}$CO, and thermal profile; all other absorbers at the prior floor), injected with $\alpha=0.90$; each injection is retrieved with the benchmark setup. For $(-250,-80)$, the \ce{H2O} and CO posteriors reach the upper prior edge; we list their medians, from which the deviations are computed. Deviations are measured in units of the posterior; a calibrated posterior has an rms deviation of one.}
\end{table*}

\section{Full comparison with \citet{Smith2024}}
\label{app:comparison}

Table~\ref{tab:smith_full} lists every retrieved parameter of the three WASP-77A\,b retrievals, including the shape parameters of the temperature profile, next to the values of \citet{Smith2024}. It also lists the temperature at three fixed pressures, which is independent of the parameterization (Sect.~\ref{sec:disc_codes}). We classify each \bear posterior with the convention of \citet{Smith2024}, which we implement as follows. A parameter is bounded, and quoted as a median with a $1\sigma$ interval, if its central $2\sigma$ interval is separated from both prior edges by more than 5\% of the prior width. If the interval reaches one edge, we quote the $3\sigma$ limit, and if it reaches both edges, the parameter is unconstrained.

\begin{table*}[htb]
\centering
\caption{All retrieved and derived parameters of the WASP-77A\,b retrievals compared with \citet{Smith2024}.}
\label{tab:smith_full}
\footnotesize
\begin{tabular}{lcccccc}
\toprule
& \multicolumn{2}{c}{IGRINS (HR)} & \multicolumn{2}{c}{NIRSpec (LR)} & \multicolumn{2}{c}{Joint (HR$+$LR)} \\
\cmidrule(lr){2-3}\cmidrule(lr){4-5}\cmidrule(lr){6-7}
Parameter & \citet{Smith2024} & \bear & \citet{Smith2024} & \bear & \citet{Smith2024} & \bear \\
\midrule
\multicolumn{7}{l}{\emph{Composition}} \\[2pt]
$\log_{10} X_\mathrm{H_2O}$ & $-3.97^{+0.11}_{-0.10}$ & $-3.74^{+0.14}_{-0.11}$ & $-3.80^{+0.34}_{-0.28}$ & $-3.90^{+0.28}_{-0.20}$ & $-4.02^{+0.07}_{-0.06}$ & $-3.69^{+0.15}_{-0.12}$ \\[3pt]
$\log_{10} X_\mathrm{CO}$ & $-3.81^{+0.19}_{-0.17}$ & $-3.65^{+0.20}_{-0.18}$ & $-3.73^{+0.55}_{-0.52}$ & $-4.16^{+0.48}_{-0.37}$ & $-3.91^{+0.13}_{-0.13}$ & $-3.76^{+0.22}_{-0.19}$ \\[3pt]
$\log_{10} X_\mathrm{CH_4}$ & $<-5.97$ & $<-5.78$ & --- & --- & $<-6.93$ & $<-6.51$ \\[3pt]
$\log_{10} X_\mathrm{H_2S}$ & $<-4.45$ & $<-3.94$ & --- & --- & $<-5.07$ & $<-4.69$ \\[3pt]
$\log_{10} X_\mathrm{NH_3}$ & $<-5.88$ & $<-5.64$ & --- & --- & $<-6.03$ & $<-5.55$ \\[3pt]
$\log_{10} X_\mathrm{HCN}$ & $<-5.69$ & $<-5.18$ & --- & --- & $<-6.12$ & $<-5.74$ \\[3pt]
$\log_{10} X_\mathrm{CO_2}$ & --- & $<-4.41$ & $<-6.34$ & $<-6.68$ & $<-7.25$ & $-7.58^{+0.28}_{-0.57}$ \\[3pt]
$[\mathrm{^{13}CO/^{12}CO}]$ & $0.63^{+0.30}_{-0.34}$ & $0.83^{+0.27}_{-0.30}$ & $<1.78$ & $<2.41$ & $0.33^{+0.35}_{-0.71}$ & $0.60^{+0.33}_{-0.49}$ \\
\midrule
\multicolumn{7}{l}{\emph{Temperature profile}} \\[2pt]
$T_0$ (K) & $1470^{+210}_{-370}$ & $1488^{+234}_{-480}$ & $1300^{+80}_{-50}$ & $1237^{+50}_{-49}$ & $1400^{+30}_{-40}$ & $1275^{+25}_{-26}$ \\[3pt]
$\log_{10} P_1$ (bar) & $>-4.25$ & $>-1.01$ & $>-1.37$ & $>-0.60$ & $0.69^{+0.22}_{-0.24}$ & $-0.16^{+0.37}_{-0.22}$ \\[3pt]
$\log_{10} P_2$ (bar) & Unc. & Unc. & Unc. & Unc. & Unc. & Unc. \\[3pt]
$\log_{10} P_3$ (bar) & $>-0.83$ & Unc. & $<-1.16$ & Unc. & $>-0.51$ & Unc. \\[3pt]
$\alpha_1$ & $>0.31$ & $>0.31$ & $0.48^{+0.04}_{-0.03}$ & $0.46^{+0.01}_{-0.01}$ & $0.60^{+0.04}_{-0.05}$ & $0.46^{+0.01}_{-0.01}$ \\[3pt]
$\alpha_2$ & Unc. & $>0.07$ & Unc. & $>0.05$ & $>0.03$ & $>0.06$ \\[3pt]
$T(0.01\,\mathrm{bar})$ (K) & --- & $1686^{+93}_{-137}$ & --- & $1638^{+53}_{-42}$ & --- & $1683^{+26}_{-23}$ \\[3pt]
$T(0.1\,\mathrm{bar})$ (K) & --- & $1816^{+86}_{-71}$ & --- & $1864^{+57}_{-39}$ & --- & $1911^{+32}_{-27}$ \\[3pt]
$T(1\,\mathrm{bar})$ (K) & --- & $1943^{+221}_{-146}$ & --- & $2137^{+59}_{-38}$ & --- & $2129^{+38}_{-31}$ \\
\midrule
\multicolumn{7}{l}{\emph{Velocities and scale factors}} \\[2pt]
$\delta\Kp$ (km\,s$^{-1}$) & $-1.26^{+0.85}_{-0.85}$ & $-0.45^{+0.90}_{-0.92}$ & --- & --- & $-1.29^{+0.75}_{-0.78}$ & $-0.34^{+0.89}_{-0.91}$ \\[3pt]
$\delta\Vsys$ (km\,s$^{-1}$) & $-5.27^{+0.51}_{-0.53}$ & $-6.60^{+0.60}_{-0.57}$ & --- & --- & $-5.27^{+0.48}_{-0.47}$ & $-6.66^{+0.60}_{-0.57}$ \\[3pt]
$\log_{10}\alpha$ (IGRINS) & $0.15^{+0.52}_{-0.53}$ & $0.25^{+0.74}_{-0.38}$ & --- & --- & $0.03^{+0.04}_{-0.04}$ & $-0.05^{+0.02}_{-0.02}$ \\[3pt]
$\log_{10}a_4$ (NIRSpec) & --- & --- & $0.01^{+0.02}_{-0.03}$ & fixed, 0 & $0.03^{+0.01}_{-0.02}$ & fixed, 0 \\
\midrule
\multicolumn{7}{l}{\emph{Derived}} \\[2pt]
C/O & $0.59^{+0.07}_{-0.07}$ & $0.57\pm0.07$ & $0.54^{+0.15}_{-0.16}$ & $0.36^{+0.13}_{-0.10}$ & $0.57^{+0.06}_{-0.06}$ & $0.47\pm0.06$ \\[3pt]
$[\mathrm{(C+O)/H}]$ & $-0.53^{+0.16}_{-0.13}$ & $-0.29^{+0.17}_{-0.15}$ & $-0.42^{+0.49}_{-0.42}$ & $-0.69^{+0.39}_{-0.27}$ & $-0.61^{+0.11}_{-0.10}$ & $-0.36^{+0.19}_{-0.16}$ \\
\midrule
\multicolumn{7}{l}{\emph{Sampling (\bear)}} \\[2pt]
Free parameters & & 17 & & 10 & & 17 \\[2pt]
Live points & & 800 & & 2500 & & 1000 \\[2pt]
Likelihood evaluations & & 573\,014 & & 183\,736 & & 625\,630 \\[2pt]
Wall time (h) & & 0.70 & & 0.09 & & 0.94 \\[2pt]
$\ln Z$ & & 381.89 & & $-892.20$ & & $-504.51$ \\
\bottomrule
\end{tabular}
\tablefoot{Values with uncertainties are medians with $1\sigma$ intervals; $<$ and $>$ are $3\sigma$ upper and lower limits; ``Unc.'' marks an unconstrained parameter and ``---'' one not included in the retrieval. The values of \citet{Smith2024} are from their pre-eclipse, NIRSpec, and pre-eclipse$+$NIRSpec retrievals, in which $a_2$ and $a_4$ are the scale factors of the pre-eclipse IGRINS and the NIRSpec data, and $\alpha$ corresponds to $a_2$. $T(P)$ is the posterior of the temperature at pressure $P$, which \citet{Smith2024} do not tabulate. $\ln Z$ is the \multinest nested-sampling evidence. The high-resolution term omits terms that depend only on the data (Sect.~\ref{sec:like}), so absolute values of $\ln Z$ have no meaning and only differences between retrievals of the same data should be compared. }
\end{table*}

\end{appendix}

\end{document}